# Beyond the Federer paradox in tennis

Andrés Martín[a], Johann H. Martínez[a] and Javier M. Buldú[a,*]

[a]Complex Systems Group, King Juan Carlos University, Calle Tulipán s/n, 28933 Móstoles, Madrid, Spain

[*]Corresponding author: Javier M. Buldú, javier.buldu@urjc.es

## Abstract

Roger Federer won about 80% of his professional matches with only 54% of the points. Hierarchical scoring amplifies small advantages on points into large advantages on matches, but also allows a player to win with fewer points than the opponent. Using Monte Carlo simulations and exact Markov chain calculations, we show that in tennis, badminton, volleyball and table tennis this amplification is set almost entirely by the number of points per match. We then let a player win ordinary points with probability $p_1$ and crucial points, those closing a game, set or match, with probability $p_2$. In every sport a narrow region of the ($p_1$, $p_2$) plane exists where a player wins most matches with fewer than half of the points, because crucial points carry more influence than their share of points. Of ATP matches played in 1991–2025, 4.7% were won with fewer points, 16% more than independent points predict (5% including the serve), a constant excess. Pressure performance measured in other seasons predicts match outcomes beyond the points won, and Fabio Fognini won 51% of his matches with 49% of the points, in what we call the Fognini paradox.

## Introduction

In his commencement address at Dartmouth College in June 2024, Roger Federer said that he had won almost 80% of the 1,526 singles matches of his professional career, but only 54% of the points (Federer, 2024). His official record, 1,251 wins (ATP Tour, 2022), corresponds to about 82% of those matches. The figure illustrates a basic property of hierarchical scoring. A tennis match is won by winning sets, sets by winning games and games by winning points, so a modest advantage on points is magnified at each level, a fact recently revisited in connection with Federer's remark (Bauer & Bauer, 2025; Lee, 2025).

Assuming that points are independent and identically distributed (i.i.d.), closed form expressions link the probability of winning a point to that of winning a game, a set and a match in tennis (Carter & Crews, 1974; Fischer, 1980; Hsi & Burych, 1971; Pollard, 1983; Riddle, 1988), and these models have been compared with data and used for prediction (Barnett & Clarke, 2005; Newton & Aslam, 2006; Newton & Keller, 2005; O'Malley, 2008). The efficiency of scoring systems has been analysed in general (Miles, 1984; Peña et al., 2026) and for table tennis (Dominicy et al., 2013; Schulman & Hamdan, 1977), badminton (Percy, 2009) and volleyball (Ferrante & Fonseca, 2014), and the predictability of competitions has been characterised across sports (Ben-Naim et al., 2006; Merritt & Clauset, 2014).

Points in professional tennis are not exactly i.i.d., although the deviations are small (Klaassen & Magnus, 2001), and the probability of winning a point depends on the score (Sim & Choi, 2020; Yoshizawa et al., 2026). Morris (1977) defined the importance of a point as the difference between the probabilities of winning the match after winning and after losing it, and showed that a player who redistributes his performance towards important points increases his chance of winning, an idea explored since (Klaassen & Magnus, 2014; Pollard, 2004). Studies of individual points show that performance changes under pressure (Cohen-Zada et al., 2017; González-Díaz et al., 2012; Knight & O'Donoghue, 2012; Kovalchik & Reid, 2018; O'Donoghue, 2001; Paserman, 2023) and that better players perform relatively better when it matters most (González-Díaz et al., 2012), but performance on break points shows little persistence from year to year (Latimer, 2022; Sackmann, 2019). A match is also sometimes won by the player who wins fewer points or games (Brams et al., 2025; Lisi et al., 2019;

Sackmann, 2015; Wright et al., 2013), a quasi-Simpson's paradox of best-of-*N* contests (Wright et al., 2013).

Federer himself is a striking example. Between 1991 and 2025, he won more points than his opponent in 1,197 ATP matches with point statistics and lost 36 of them (3.0%), more than any other player. He was the loser in 36 of the 45 paradoxical matches he played (80.0%), a record first reported by Wright et al. (2013) and called the Federer paradox by Lisi et al. (2019). Read the other way round, however, of the 232 matches in which Federer won fewer points than his opponent, he won 9 (3.9%), a higher share than the 3.0% he lost after winning more points. A dominant player simply has many more opportunities to lose a paradoxical match than to win one, so his record is expected (Lisi et al., 2019) and can even be read as evidence of his greatness (Wright et al., 2013). We show below that it is partly a counting effect and partly a small but genuine deficit.

What has not been established is how a difference in performance between ordinary and crucial points maps onto the probability of winning a match, and when it can make a player who wins fewer than half of the points the favourite. We address this question with a minimal model in which a player wins ordinary points with probability $p_1$ and crucial points, those whose winner closes a game, a set or the match, with probability $p_2$. We compute the probability of winning exactly for tennis, badminton, volleyball and table tennis, characterise the region where a player wins most matches with fewer than half of the points, and confront the model with three decades of men's professional tennis, extending previous analyses of points and wins (Bauer & Bauer, 2025) and of performance under pressure (González-Díaz et al., 2012; Sackmann, 2019) with point by point data, out of sample tests and the history of the paradox.

## Methods

### Scoring systems

In tennis, a game is won by the first player to win four points with a margin of two, a set by the first to win six games with a margin of two, with a tiebreak to seven points at six games all, and the match by the first to win two or three sets. Badminton (games to 21 points, capped at 30, best of three), volleyball (sets to 25 points, deciding fifth set to 15, best of five) and table tennis (games to 11 points, best of five or seven) use rally scoring with a margin of two. We use the term *set* for the scoring unit immediately above the point in these sports (Supplementary Table S2). The serve is not modelled, which isolates the effect of crucial points. Adding it with the rules of each sport changes the amplification by at most 8% and does not change the conclusions (Supplementary Section S12).

**Ordinary and crucial points**

Player A wins each point independently, with probability $p_1$ if the point is ordinary and $p_2$ if it is crucial. A point is crucial at level $L \in$ {game, set, match} if the player who wins it, whichever of the two, closes a scoring unit of level $L$ or higher. At the game level every game point is crucial, including break points, and points at deuce are ordinary. Crucial points at the set and match levels are the subsets of game points that would also close a set or the match. With probability $p_2$ player A therefore converts his own game, set and match points and saves those of the opponent. In a variant in which only the points that A can close are played with probability $p_2$, the paradox region still exists, with 40% to 50% of the opening angle (Supplementary Section S5). For $p_1 = p_2$ the model reduces to the classical i.i.d. model.

**Simulations and exact calculation**

A match is played between players A and B. $N_A$ and $N_B$ are the points won by each in a match, $P(p_1, p_2)$ is the probability that A wins the match, and $F(p_1, p_2) = E[N_A]/(E[N_A] + E[N_B])$, where $E[\cdot]$ denotes the expected value, is the share of points won by A over many matches, a ratio of totals like a share of points over a career. $N = E[N_A + N_B]$ at $p_1 = p_2 = 1/2$ is the mean number of points per match between two equal players. Matches were simulated point by point (20,000 matches per value of $p$ for a single probability, and 2,000 per point of a grid of step 0.02 in the ($p_1$, $p_2$) plane). Because the outcome of each point depends only on the score, the score is a Markov chain, and $P$ and $F$ were also computed exactly by dynamic programming, on a grid of step 0.01 and on finer grids around (1/2, 1/2). The exact calculation reproduces the classical probability of winning a tennis game (Carter & Crews, 1974), and the simulations agree with it within sampling error (Supplementary Section S4). The algorithms, the partial derivatives at (1/2, 1/2) and the derivation of Eq. (4) are given in Supplementary Section S13.

**Data**

We used two databases compiled by Sackmann (n.d.) and released under a CC BY-NC-SA 4.0 licence, from an archival copy made in June 2026 (tennis-sackmann-archive, 2026) after the original repositories were withdrawn. The ATP match database gives, for each match of the main tour, aggregate statistics (serve points played and won by each player and break points faced and saved), available from 1991. After removing retirements, walkovers and matches without statistics, it contains 97,950 matches of the seasons 1991–2025 and 2,559 players, and all analyses of players and seasons use it. The Grand Slam point by point database records the winner, the server and the score after every point for 5,336 men's singles matches of 49 tournaments between 2011 and 2024. We reconstructed the score before each point and classified every point as in the model, taking into account the final set rules of each tournament and year. Every point that ended a game or a set was crucial at that level for its winner in 99.92% of games and 99.44% of sets. We excluded the Australian Open and Roland Garros files of 2018–2021 (1,012 matches), in which the order of points is scrambled, and 262 incomplete or inconsistent matches (Supplementary Section S7), leaving 4,062 matches, used only where the score of every point is needed.

**Measures of performance**

We use three measures of how well a player turns points into wins.

**Match conversion.** A player's share of matches is mostly determined by his share of points (Fig. 4a), and the i.i.d. model at his career share of points is a poor benchmark, because that share averages over opponents of very different strength. We therefore compare every match with its own benchmark. The *match conversion* of a player is

$$C = (1/n)\ \Sigma_{m}\ [w_{m} - \pi(f_{m})], \qquad (1)$$

where the sum runs over his $n$ matches, $w_m = 1$ if he won match $m$ and 0 otherwise, $f_m$ is the share of points he won in it, and $\pi(f)$ is the probability of winning a match with a share of points $f$ when all points are i.i.d., estimated from $4 \times 10^6$ simulated matches of each format with $p$ uniformly distributed in [0.3, 0.7] ($\pi = 1/2$ when both players won the same number of points, and $\pi(1 - f) = 1 - \pi(f)$). The strength of the opponent enters only through $f_m$. $C$ is given per 100 matches, and a player who wins crucial points more often than ordinary ones has $C > 0$.

**Pressure index.** The ATP statistics record only one kind of crucial point, the break point, so in these data the comparison of $p_2$ and $p_1$ is restricted to break points. The raw difference between the shares of break points and of other points won is misleading, because a strong server plays most of his break points on return, where he wins few points, and a weak server on serve. The *pressure index* extends the break points over expectation of Sackmann (2019) to the break points defended on serve. Unlike indices that rate the importance of each game or point (Cohen-Zada et al., 2017; Harris et al., 2021), it is a property of the player, and it compares every break point with his level on the same kind of point:

$$I = [B - (b_s r_s + b_r r_r)]/(b_s + b_r), \qquad (2)$$

where $B$ is the number of break points he won, $b_s$ and $b_r$ are the break points he played on serve and on return, and $r_s$ and $r_r$ are the shares of the other points he won on serve and on return. $I > 0$ means that he plays break points better than his other points, and $I$ is given in percentage points. In the Grand Slam data the index uses all the crucial points of the model.

**Close match balance.** The numbers of matches that a player won with fewer than half of the points, $W$, and lost with more than half, $L$, depend on how often he wins more points than his opponent. The *close match balance* subtracts what independent points would give:

$$\Delta = (W - L) - (E[W] - E[L]), \qquad (3)$$

where $E[W]$ is the sum of $\pi(f_m)$ over the matches with $f_m < 1/2$ and $E[L]$ the sum of $1 - \pi(f_m)$ over those with $f_m > 1/2$, with Poisson-binomial variance $\Sigma\, \pi(1 - \pi)$. $\Delta$ is measured in matches, and it equals $nC$ computed without the matches in which both players won the same number of points.

**Set measures.** From the scores we computed the share of close sets won (decided by two games or fewer), the mean margin of the sets won and lost, their difference (the *set margin asymmetry*, positive when lost sets are more lopsided) and the share of lost sets that ended 6–0 or 6–1.

**Statistical analysis**

The expected number of matches won with fewer than half of the points is the sum over matches of $\pi(f_{low})$, where $f_{low}$ is the smaller share of points, with Poisson-binomial variance, from which $z$-scores and confidence intervals are obtained. Temporal trends were tested with logistic regressions of the indicator for each match on the season (linear and quadratic terms), with and without the logit of $\pi(f_{low})$ as a covariate, and with weighted least squares on the yearly excess. For the out of sample test, the pressure index of each player in each season was computed from all his *other* seasons (players with at least 200 break points in them), and the outcome of each match, oriented towards a randomly chosen player, was regressed on the logit of $\pi(f)$ and on the difference between the two players' indices. The stability of each index was assessed by correlating it between odd and even seasons, and its standard deviation across players was corrected by subtracting the mean binomial variance. Confidence intervals of player measures come from 2,000 bootstrap resamples of matches, $P$-values were corrected with the Benjamini–Hochberg procedure, and correlations are reported with Fisher-$z$ intervals.

The tiebreak overperformance is the difference between the tiebreaks a player won and those expected at the share of points he won in each match, divided by the number of tiebreaks. The games paradox was computed from all completed main tour matches of 1968–2025. The three channels of the close match balance were compared with a least squares regression on standardised variables. To test the robustness of $\pi$ to the serve, it was also estimated from matches simulated with the serve, conditioning on the serve dominance of each match (Supplementary Section S11).

## Results

### Amplification of a point advantage

We first consider the classical case in which player A wins every point with the same probability $p$. The probability $P(p)$ of winning a match passes through $P(1/2) = 1/2$ and becomes steeper as matches get longer (Fig. 1a). The slope at $p = 1/2$, which measures how strongly a small advantage is amplified, ranges from 7.3 in table tennis over five games to 13.4 in tennis over five sets (Supplementary Table S1). A player who wins 52.5% of the points wins 80% of his matches in tennis over five sets, whereas in table tennis over five games he needs 54.6% of the points. The slope is well approximated by

$$(\mathrm{d}P/\mathrm{d}p)_{\mathrm{p}\,=\,1/2} \approx (2N/\pi)^{1/2}, \qquad (4)$$

where $N$ is the mean number of points per match at $p = 1/2$, with deviations of at most 5% for all six formats (Fig. 1b). Equation (4) is the slope of the simplest contest, in which $N$ points are played and the player who wins more of them wins. A point decides the outcome only when the other points are split evenly, and the probability of this falls as $N^{-1/2}$ (Supplementary Section S13). The amplification is therefore set almost entirely by the length of the match, and the hierarchy of games and sets modifies it only slightly (+5% in tennis over three sets, −3% in volleyball).

For $p = 0.54$, the i.i.d. model predicts that a player wins 86% of matches over three sets and 91% over five sets, more than Federer's 82%. Federer's 54% is an average over opponents of very different strength, and because $P(p)$ is concave above 1/2, the average of $P$ is smaller than $P$ at the average $p$ (Jensen's inequality; Jensen, 1906).

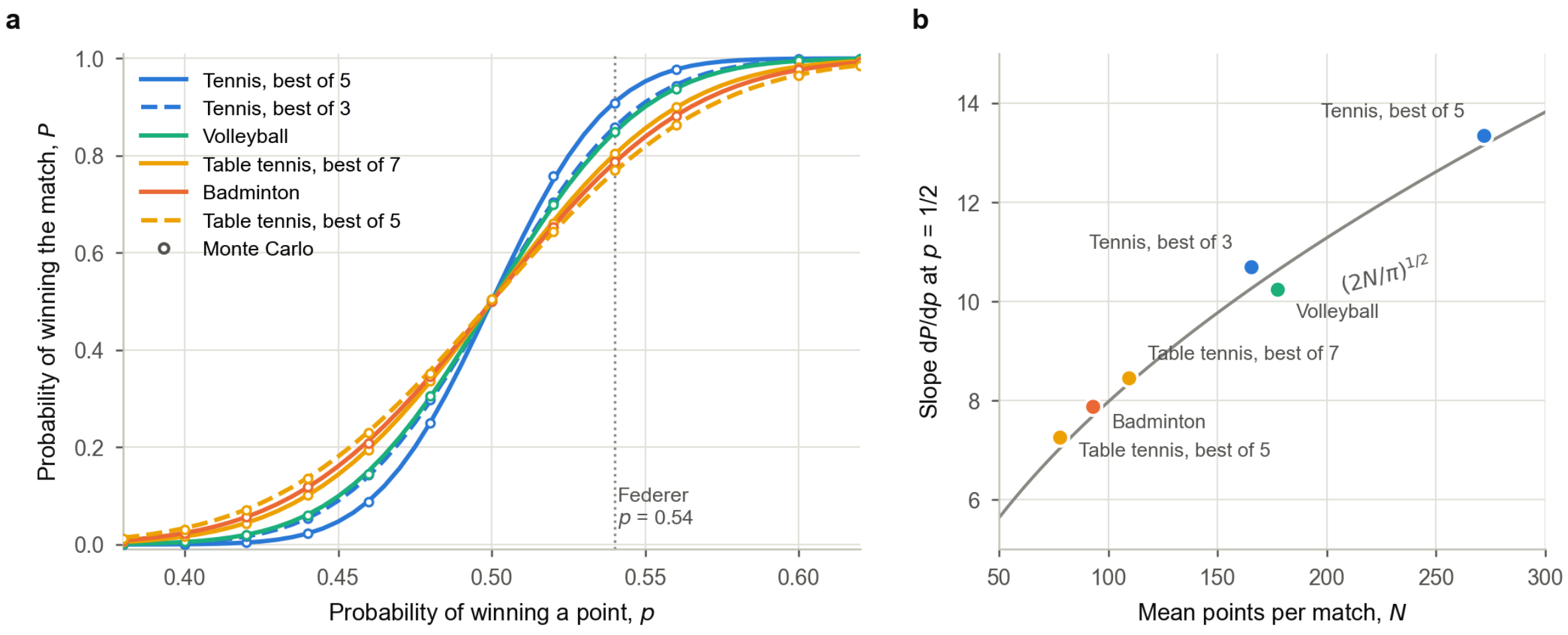

**Figure 1. From points to matches.** (a) Probability $P$ of winning a match against the probability $p$ of winning a point, for tennis over five and three sets, volleyball, table tennis over seven and five games and badminton. Lines, exact calculation. Symbols, Monte Carlo simulations (20,000 matches per value of $p$). Dotted line, $p = 0.54$. (b) Slope $dP/dp$ at $p = 1/2$ against the mean number of points per match $N$, with the estimate $(2N/\pi)^{1/2}$ of Eq. (4) (line).

## Crucial points and the paradox region

We now let player A win ordinary points with probability $p_1$ and crucial points with probability $p_2$. Figure 2 shows the probability of winning a tennis match in the $(p_1, p_2)$ plane. Game points are frequent, about 29% of all points at $p_1 = p_2 = 1/2$, so $p_2$ at the game level controls a large part of the match, whereas set and match points, about 3% and 1% of the points, matter only when $p_2$ differs strongly from $p_1$.

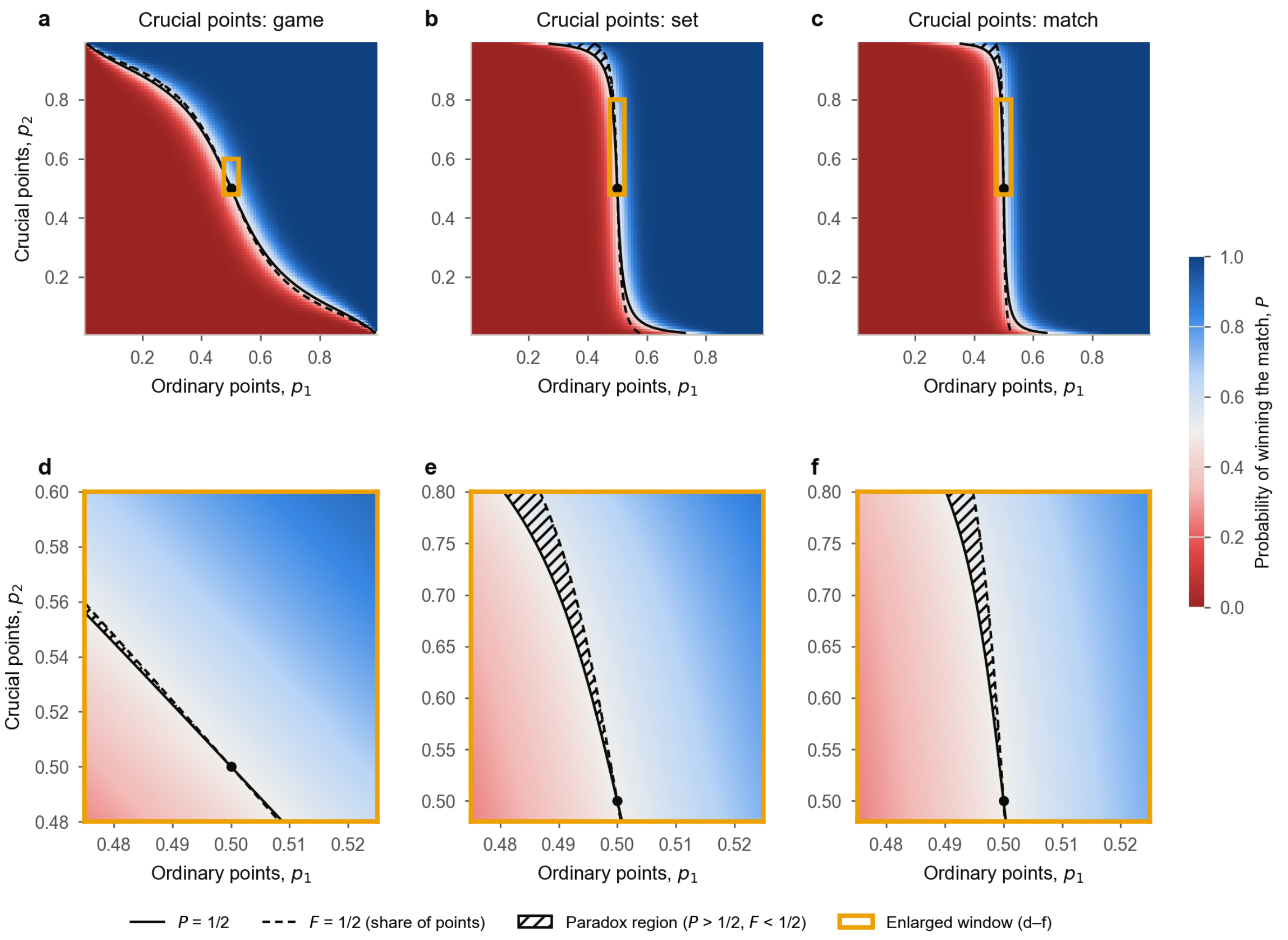

**Figure 2. Crucial points in tennis.** Probability $P$ of winning a match over three sets (colour) when player A wins ordinary points with probability $p_1$ and crucial points with probability $p_2$, with crucial points at the level of (a) the game, (b) the set and (c) the match. Solid lines, $P = 1/2$. Dashed lines, $F = 1/2$. Hatched area, paradox region ($P > 1/2$ and $F < 1/2$). (d–f) Enlargement of the orange rectangles of panels a–c, for $p_1$ between 0.475 and 0.525. Exact calculation.

Let $F(p_1, p_2)$ be the expected share of points won by A, the analogue of Federer's 54%. We call *paradox region* the set of $(p_1, p_2)$ for which $P > 1/2$ and $F < 1/2$, where player A wins most of his matches while winning fewer than half of the points. It exists for every format and every level of crucial point. Both boundaries, $P = 1/2$ and $F = 1/2$, pass through (1/2, 1/2), where the two players are equivalent. Linearising around this point,

$$P - 1/2 \approx P_1\delta_1 + P_2\delta_2, \quad F - 1/2 \approx F_1\delta_1 + F_2\delta_2, \tag{5}$$

where $\delta_i = p_i - 1/2$ and $\{P_i, F_i\}$ are the partial derivatives at (1/2, 1/2). A point with $\delta_1 < 0 < \delta_2$ belongs to the paradox region if

$$P_2/P_1 > -\delta_1/\delta_2 > F_2/F_1. \tag{6}$$

The paradox region is therefore a wedge with its vertex at (1/2, 1/2), which exists if and only if $P_2/P_1 > F_2/F_1$. The ratio $F_2/F_1$ is essentially the ratio between the numbers of crucial and ordinary points, whereas $P_2/P_1$ is the ratio of their influence on the outcome. Crucial points must therefore carry a larger share of the influence than their share of the points, which is Morris' notion of point importance (Morris, 1977), and the condition holds for every sport and level (Supplementary Table S3).

The region is narrow, with opening angles from 0.3° to 1.6° (Fig. 3b and Supplementary Table S3). In tennis at the game level, a player with $p_1 = 0.49$ wins most matches with fewer than half of the points only if $p_2 \approx 0.524$, and with $p_1 = 0.48$ only if $p_2 \approx 0.546$, because any advantage on frequent game points also raises the share of points. At the set and match levels crucial points are rare and barely change the share of points, but an advantage of $p_2 \gtrsim 0.7$ is needed to tip the match. Table tennis, with many set points, has the widest wedge (Fig. 3).

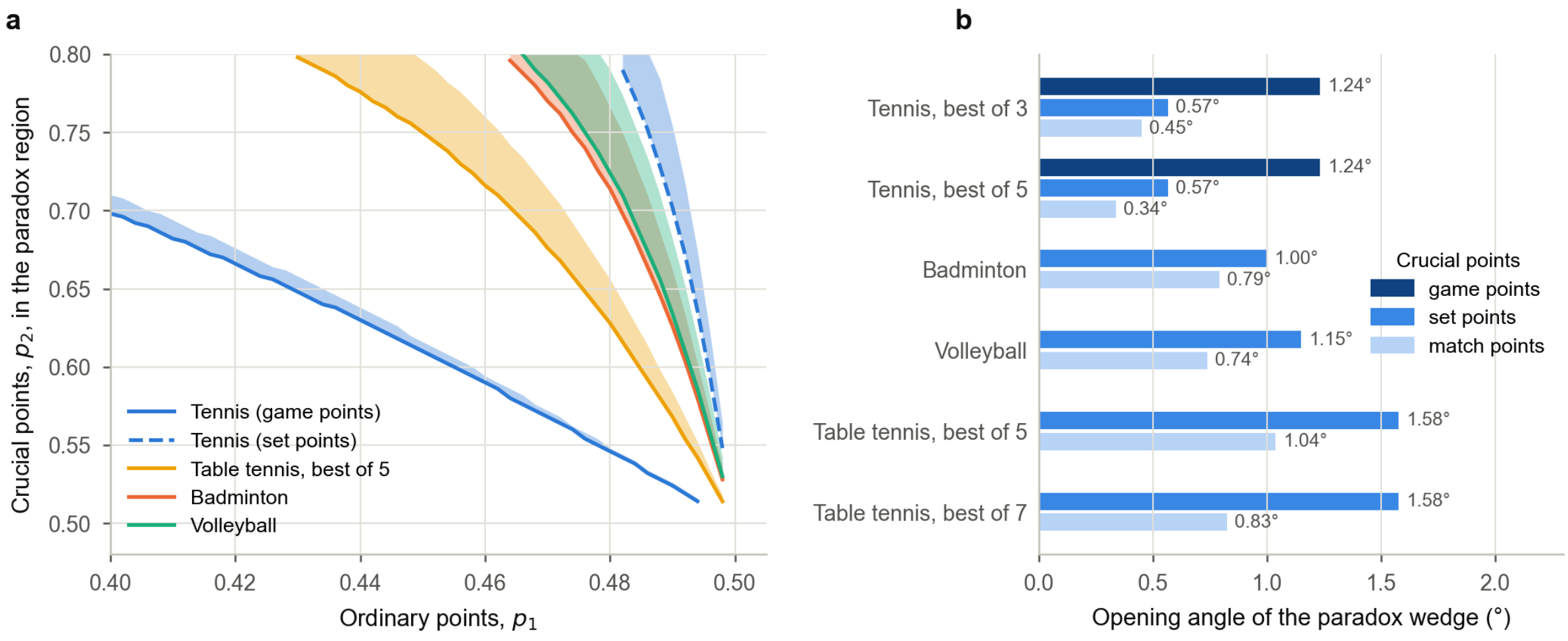

**Figure 3. Paradox regions across sports.** (a) Range of $p_2$ for which a player with probability $p_1$ on ordinary points wins most matches with fewer than half of the points (shaded bands), for tennis with crucial points at the game and set levels and for table tennis (best of five), badminton and volleyball at the set level. (b) Opening angle of the paradox wedge for every format and level. Exact calculation.

## Points and wins in professional tennis

Across the 328 players with at least 200 ATP matches between 1991 and 2025, the shares of points and of matches won are almost perfectly correlated (Pearson $r = 0.97$, 95% confidence interval (CI) 0.965–0.977, Fig. 4a), as reported for a similar data set (Bauer & Bauer, 2025). Federer won 54.1% of the points and 81.7% of his 1,437 matches with statistics. The i.i.d. model evaluated at each player's career share of points overestimates the win rate of the best players, by 5.6 percentage points on average for $p > 0.52$, a consequence of averaging over opponents. Match conversion avoids this bias and is uncorrelated with the share of points ($r = -0.04$, 95% CI −0.15 to 0.07, Fig. 4b).

Winning a match with fewer than half of the points happened in 4,530 of the 96,826 ATP matches in which the two players did not win the same number of points (4.68%), in line with the 4.5% reported for 1991–2017 (Lisi et al., 2019; Wright et al., 2013). Given the share of points won in each match, the i.i.d. model predicts 3,892 such matches (4.02%), so the observed number exceeds the expectation by 16% ($z = 11.7$, $P < 10^{-31}$, Fig. 4c–f). The excess is larger over five sets (23%, $z = 6.8$) than over three sets (15%, $z = 9.8$), in Grand Slams (22%) than in other tournaments (14% to 17%), and on grass (21%) and hard courts (19%) than on clay (13%) (Supplementary Table S7 and Fig. S5). This does not contradict the conclusion that the overall frequency is compatible with i.i.d. points (Lisi et al., 2019), which depends on the assumed differences between opponents, whereas conditioning on the points of each match removes this dependence. Part of the excess reflects the serve. With a benchmark that includes it (Supplementary Section S11), the excess falls to 5% ($z = 3.9$), and to 14% over five sets ($z = 4.5$), and the comparisons between players are essentially unchanged.

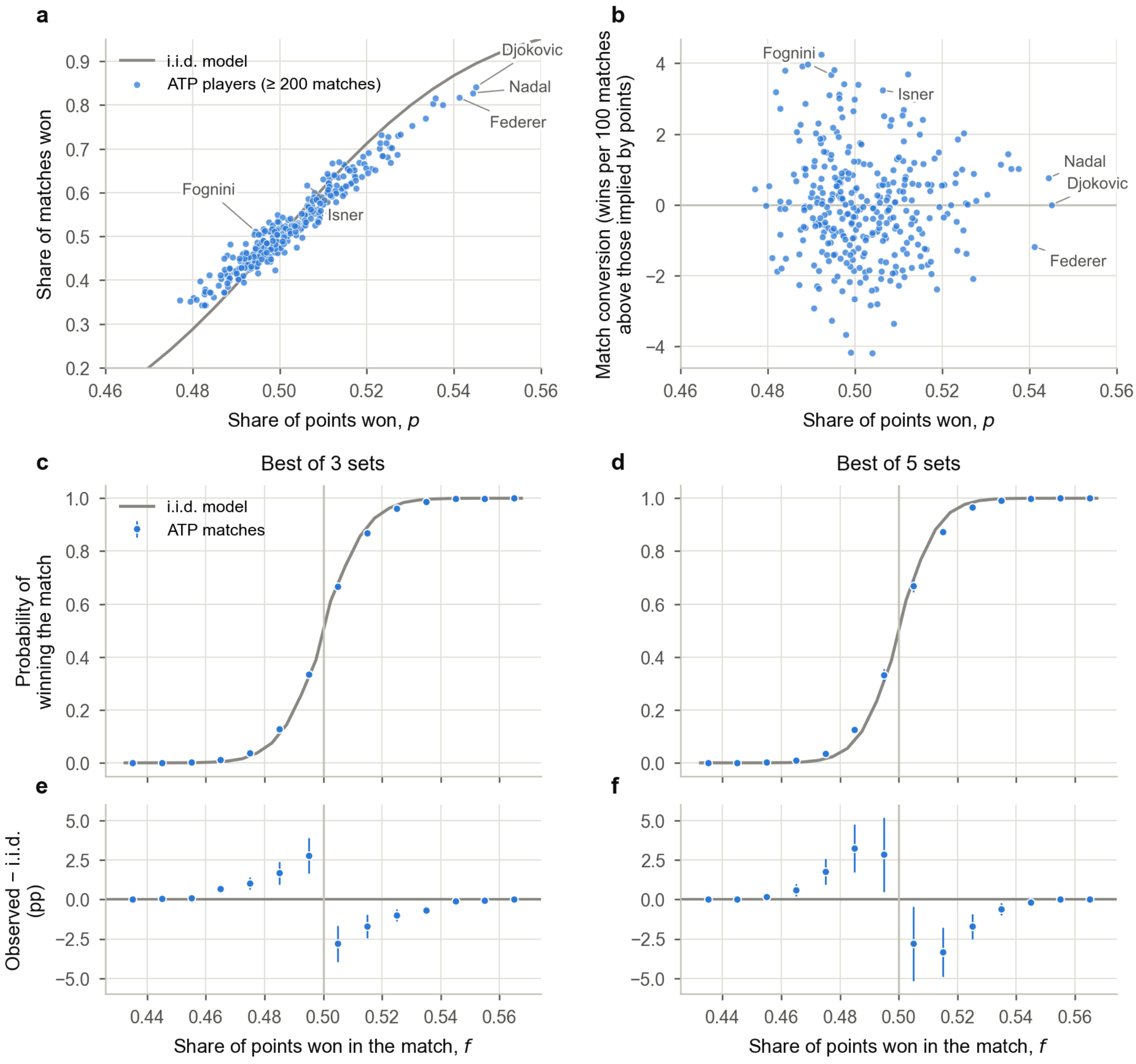

**Figure 4. Points and wins in professional tennis.** (a) Share of matches won against share of points won for the 328 players with at least 200 ATP matches (1991–2025), and the i.i.d. model for the average mix of matches over three and five sets (line). (b) Match conversion against share of points won. (c, d) Probability of winning a match as a function of the share *f* of points won in it, over three (c) and five (d) sets (symbols, bins of width 0.01), and i.i.d. model (lines). (e, f) Difference between the observed and i.i.d. probabilities. Error bars, 95% binomial confidence intervals.

The Grand Slam point by point data show where the excess comes from. In the 166 matches won with fewer than half of the points, the winners won 49.1% of the ordinary points but 51.1% of the set points and 61.9% of the match points, the mechanism of the model, $p_1 < 1/2 < p_2$, at the level of sets. At the game level the difference is small (+0.5 percentage points), consistent with the narrow paradox band at that level. These figures are partly mechanical, because the winner of a set wins the point that closes it, so the next tests measure pressure performance in other matches.

**Who benefits from crucial points?**

The pressure index varies little across players. After subtracting the binomial noise, its standard deviation across players is about 0.5 percentage points, and its correlation between odd and even seasons is only 0.13 (0.26 for players with at least 500 matches). Its effect is nevertheless clear out of sample. In a logistic regression of the outcome of 90,229 matches on the i.i.d. probability of winning given the points won and on the difference between the two players' pressure indices measured in all *other* seasons, the pressure difference has a coefficient of 0.076 per percentage point (standard error 0.011, $z = 7.2$, $P < 10^{-12}$), and 0.075 ($z = 6.5$) in close matches with $0.48 < f < 0.52$. An advantage of 1 percentage point in pressure index therefore multiplies the odds of winning by 1.08 at a given share of points. The index measured in odd seasons is also correlated with the match conversion in even seasons ($r = 0.17$, $P = 0.002$, Fig. 5c), and match conversion is a stable trait ($r = 0.35$ between odd and even seasons). The low persistence reported previously (Latimer, 2022; Sackmann, 2019) thus reflects the low reliability of the index, and pooling seasons reveals its small persistent component.

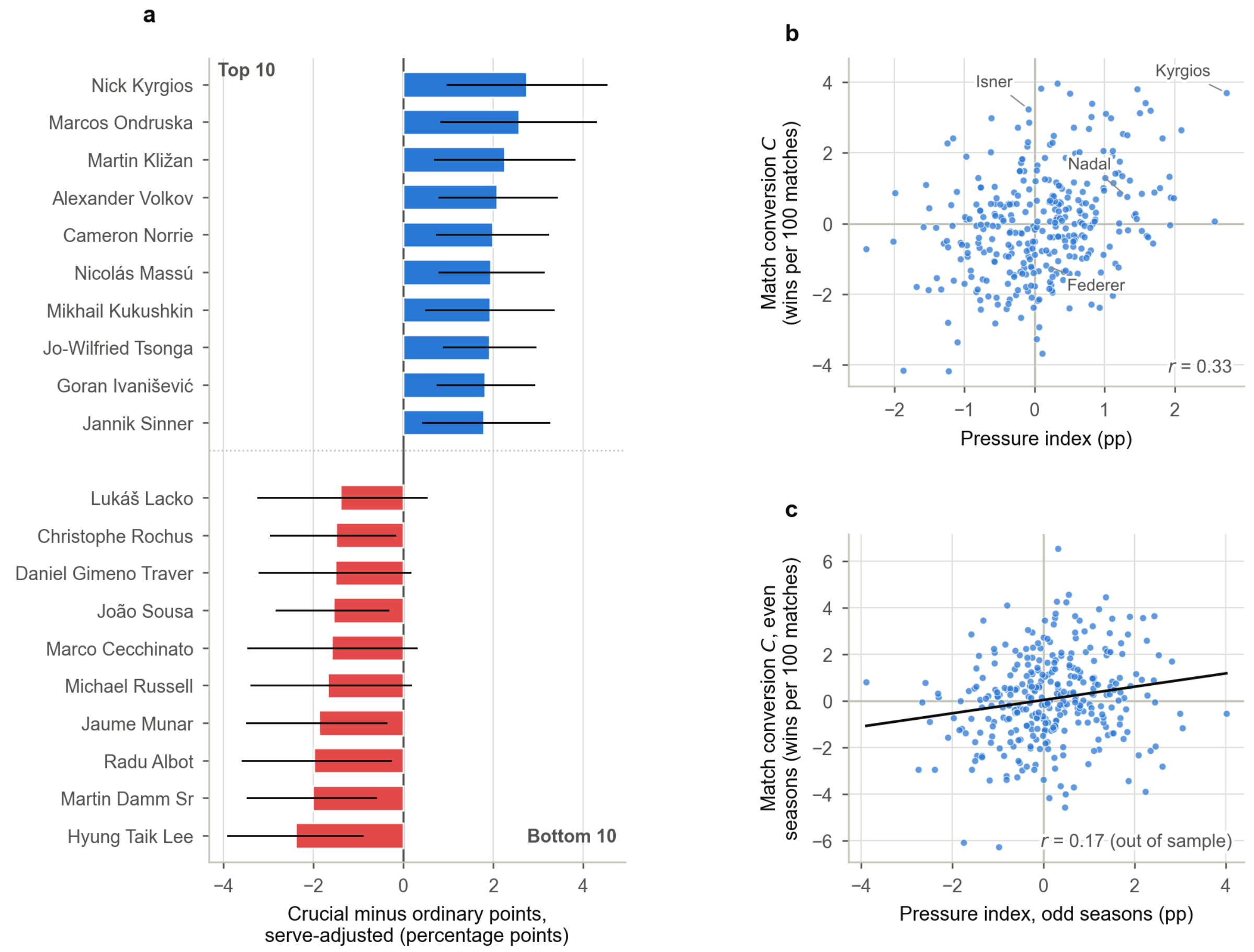

**Figure 5. Pressure and match conversion.** (a) The ten players with the largest and the ten with the smallest pressure index (at least 200 ATP matches, 1991–2025), with 95% bootstrap confidence intervals. (b) Match conversion against pressure index for the 328 players. (c) Match conversion in even seasons against pressure index in odd seasons (players with at least 60 matches in each half), with a least squares fit.

Figure 5a ranks the players by pressure index. The adjustment for the serve matters. The raw difference $p_2 - p_1$ is dominated by playing style (correlation −0.40 with the share of serve points won) and would place Federer, Alcaraz, Roddick, Raonic and Djokovic at the bottom (Supplementary Table S6). With the adjustment, Nick Kyrgios, Marcos Ondruska, Martin Klizan, Alexander Volkov and Cameron Norrie lead the ranking, although none remains significant after correcting for multiple comparisons. Across players, the pressure index and match conversion are correlated ($r$ = 0.33, Fig. 5b), and 20 of the 328 players have a match conversion significantly different from zero (false discovery rate 5%, Supplementary Table S12). Match conversion is also correlated with the overperformance in tiebreaks ($r$ = 0.45, 95%

CI 0.36–0.53), which is unrelated to the pressure index ($r$ = 0.09, $P$ = 0.12). Break points and tiebreaks are thus two distinct channels, and Isner's advantage, for instance, lies entirely in tiebreaks (Supplementary Fig. S6).

**Beyond the Federer paradox**

The close match balance (Methods) counts the close matches that a player has gained, or lost, with respect to independent points. Across the 328 players its standard deviation is 1.4 matches per 100 matches played, against 0.9 from binomial noise alone, so the genuine differences between players amount to about 1.1 matches per 100. For 23 players the balance differs significantly from zero after correcting for multiple comparisons, 16 of them positively (Fig. 6a and Table 1).

Roger Federer is the clearest negative case. His balance of paradoxical matches, 9 won and 36 lost, is the most negative in the data set, the Federer paradox (Lisi et al., 2019). Compared with the expectation, however, in the 1,197 matches in which he won more points the i.i.d. model predicts 25.6 defeats, against 36 observed ($z$ = 2.4), and in the 232 matches in which he won fewer points it predicts 15.6 victories, against 9 ($z$ = −1.9). Overall, Federer won 17 fewer close matches than his points imply ($z$ = −3.0), the same deficit measured by his match conversion, −1.2 wins per 100 matches. The paradox is therefore partly a counting effect (Lisi et al., 2019) and partly a small deficit, at the limit of significance after correcting for multiple comparisons ($q$ = 0.046) and not significant when the serve is taken into account ($z$ = −2.4, $q$ = 0.12). None of the other champions shows it. Pete Sampras (+11.2 matches, $z$ = 2.4), Andre Agassi (+8.4, $z$ = 2.0), Andy Murray (+7.8, $z$ = 1.6) and Rafael Nadal (+7.0, $z$ = 1.5) won more close matches than their points imply, and Novak Djokovic is close to the expectation (Supplementary Table S9).

At the opposite end stands Fabio Fognini. In his 774 matches he won 61,624 of 124,623 points, 49.4%, significantly fewer than half ($z$ = −3.9), yet he won 396 matches, 51.2%, whereas the

i.i.d. model predicts 47.5% given the points he won in each match. He won 32 matches with fewer points than his opponent, against 14.8 expected, and lost only 10 with more points, against 18.8 expected. His close match balance, +26.0 matches ($z$ = 5.2), is the largest of the data set, significant after correcting for multiple comparisons and robust to the serve. Fognini is a real realisation of the paradox region of the model, which we call the *Fognini paradox*. Twenty-three of the 328 players won more than half of their matches with less than half of the points (Fig. 6b and Supplementary Table S11), among them Feliciano López, Alexander Bublik and Jürgen Melzer, although mostly by small margins.

Fognini's advantage does not come from break points, where his pressure index is close to zero (+0.5 percentage points), but from the way his points are distributed across sets. He won 50.5% of all his sets but 56.6% of those decided by two games or fewer. His lost sets were more lopsided than his won sets (mean margins of 3.0 and 2.7 games), and 18% of the sets he lost ended 6–0 or 6–1, against 5% for Federer (Supplementary Fig. S7). Points lost in lopsided sets do not cost matches, as already suggested by Wright et al. (2013), who quoted Isner on saving energy in return games once a break up. Across players, the set margin asymmetry is correlated with the close match balance ($r$ = 0.33, 95% CI 0.23–0.42, Fig. 6c).

Three channels therefore turn points into wins, namely break points, tiebreaks and the distribution of points across sets. In a multiple regression across players they contribute almost equally to the close match balance, with standardised coefficients of 0.34 for the pressure index, 0.35 for the tiebreak overperformance and 0.34 for the set margin asymmetry (all $P < 10^{-11}$), and together they explain 37% of its variance.

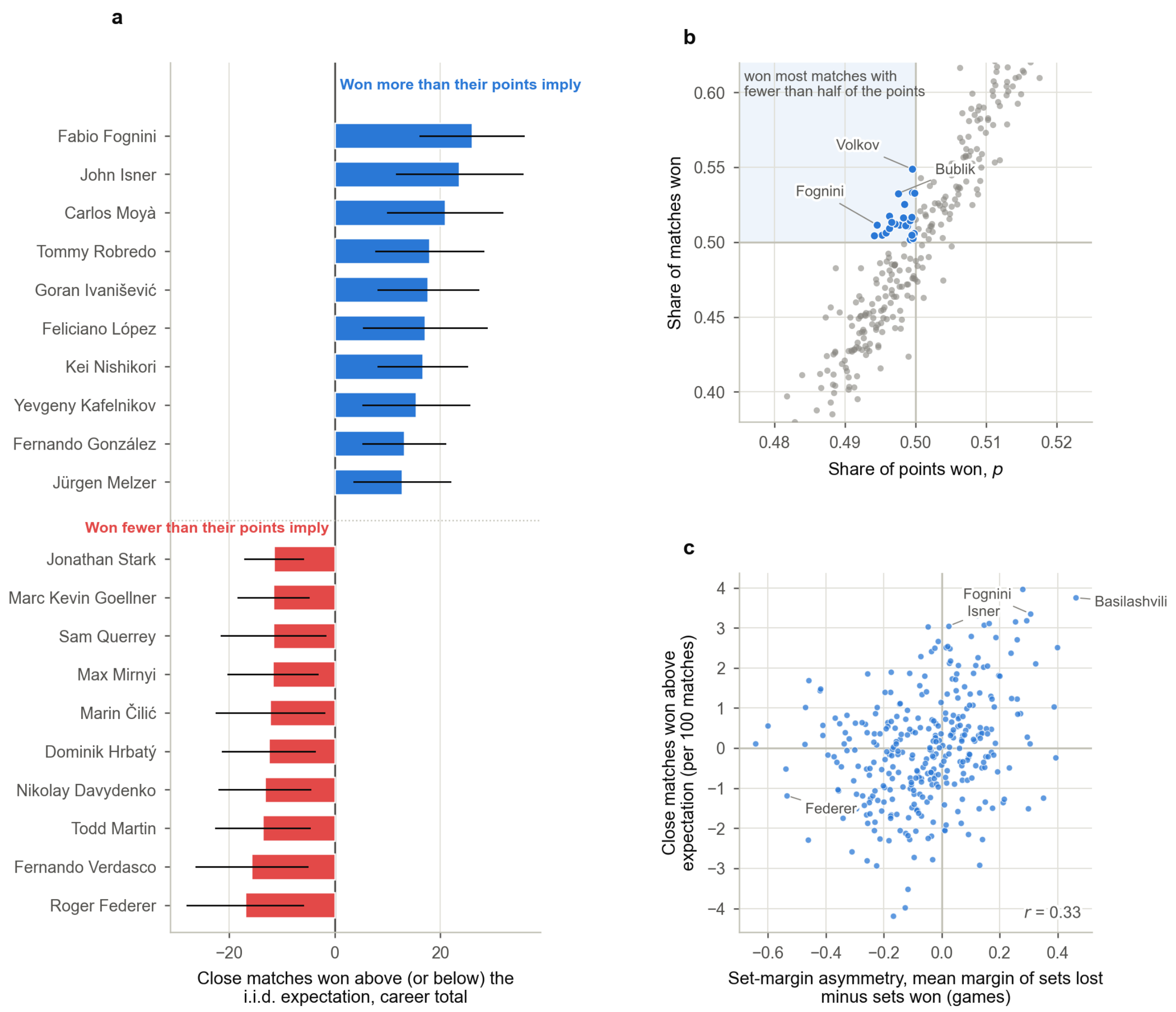

**Figure 6. Beyond the Federer paradox.** (a) The ten players with the largest and the ten with the smallest close match balance, the matches won with fewer than half of the points minus those lost with more than half, relative to the i.i.d. expectation. Lines, 95% intervals. (b) Share of matches won against share of points won around 1/2. Blue symbols, the 23 players who won more than half of their matches with less than half of the points (shaded quadrant). (c) Close match balance per 100 matches against set margin asymmetry.

## A history of the paradox

Between 1991 and 2025 the share of ATP matches won with fewer than half of the points followed a U shape (quadratic term of a logistic regression on the season, $P = 0.0003$), with a minimum around 2007 and no linear trend ($P = 0.46$, Fig. 7a). Earlier analyses found the lowest rate in the same period, but no significant differences between periods (Lisi et al., 2019; Wright et al., 2013). The excess over the i.i.d. expectation, 0.66 percentage points on average, has remained constant (trend −0.04 percentage points per decade, $P = 0.51$, Fig. 7b). The U shape reflects instead the closeness of matches, whose share fell during the 2000s and rose afterwards (Fig. 7c), and accounting for it reduces the curvature ($P$ = 0.03). Matches won with fewer games than the opponent, a paradox first analysed by Lisi et al. (2019), became slowly more frequent, from 1.64% in 1968–1990 to 1.93% in 1991–2007 and 2.00% in 2008–2025 (logistic trend 0.059 per decade, $P < 10^{-7}$, Fig. 7d), and the two paradoxes are correlated across seasons ($r$ = 0.58, $P$ = 0.0003).

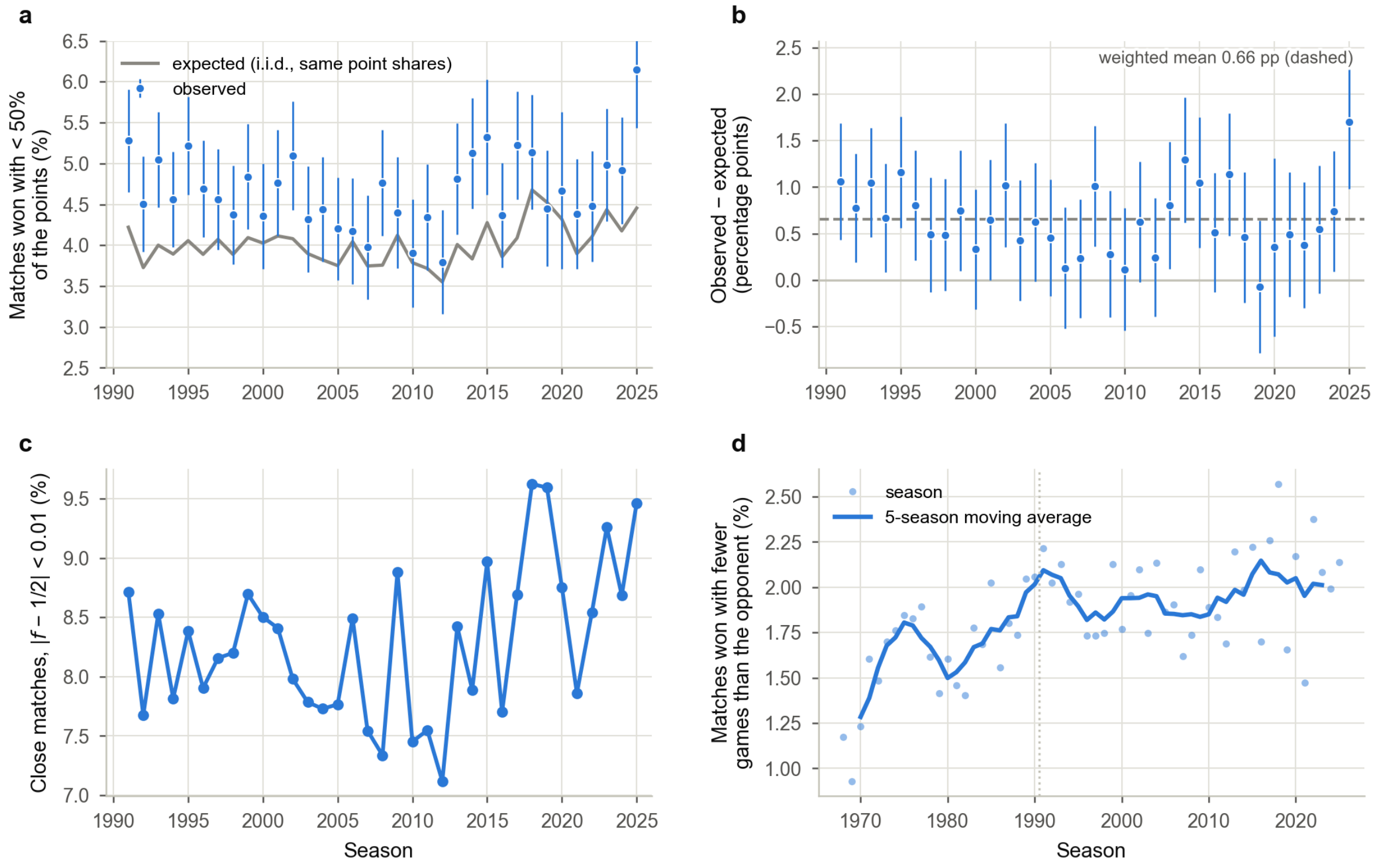

**Figure 7. A history of the paradox.** (a) Share of ATP matches won by the player with fewer than half of the points in each season, 1991–2025 (symbols, with 95% confidence intervals), and its i.i.d. expectation (line). (b) Difference between the observed and expected shares, with the weighted mean (dashed line). (c) Share of close matches, in which the shares of points differ from 1/2 by less than 0.01. (d) Share of matches won with fewer games, 1968–2025 (symbols, seasons, line, moving average over five seasons). Dotted line, start of the point statistics in 1991.

## Discussion

Federer's remark captures the strength of hierarchical scoring. We have shown that the amplification of a point advantage depends, to a very good approximation, only on the number of points played, as for a random walk, so that tennis over five sets rewards the better player most and table tennis over five games least.

The same hierarchy makes the outcome sensitive to *when* points are won. A player slightly worse on ordinary points but better on crucial points can win most of his matches with fewer than half of the points. This paradox region exists in all four sports and at every level of their scoring hierarchy, as a direct consequence of point importance, but it is narrow. Frequent crucial points cannot be won more often without raising the share of points won, and infrequent ones require a large advantage to change the outcome. This explains why matches won with fewer points occur regularly, while players who systematically win more matches than their points imply are rare.

Men's professional tennis agrees with this picture. Matches won with fewer points are 16% more frequent than an i.i.d. model predicts (5% once the serve is included), their winners perform better on set points than on ordinary points, and pressure performance measured in other seasons predicts match outcomes beyond the points won. These results agree with previous evidence on performance under pressure (Cohen-Zada et al., 2017; González-Díaz et al., 2012; Knight & O'Donoghue, 2012; O'Donoghue, 2001; Paserman, 2023) and on matches won with fewer points or games (Brams et al., 2025; Lisi et al., 2019; Sackmann, 2015; Wright et al., 2013). Our contribution is to explain why such outcomes arise, to give the condition under which they can be expected and to quantify how this depends on the scoring system. The excess due to crucial points has been constant since 1991, and the variations of the paradox follow the competitive balance of the tour. Some players, with Fognini as the clearest example, won most of their matches with fewer than half of the points by winning close sets and conceding lopsided ones, whereas Federer's paradoxical record is partly a counting effect and partly a small deficit in close matches. For performance analysts, match conversion, the pressure index and the close match balance can be computed from routine match statistics and complement the share of points won when players are evaluated.

Our model is deliberately minimal. It ignores the serve, although including it changes the amplification by at most 8% and preserves the paradox region (Supplementary Section S12). It assumes that $p_1$ and $p_2$ are constant during a match, and it treats all crucial points of a level as equivalent, whereas their importance depends on the score (Klaassen & Magnus, 2014; Morris, 1977). The data analysis is limited to men's tennis, to break points in the ATP statistics and to the Grand Slam matches recorded point by point. A continuous measure of point importance, women's tennis and other sports are natural extensions. In Federer's words, it is only a point. But not all points are equal.

## Acknowledgements

We thank Jeff Sackmann for compiling and releasing the tennis data used in this work.

## Funding

This work was supported by the Ministerio de Ciencia e Innovación, Spain, under grant PID2023-147827NB-I00 (J.M.B. and J.H.M.).

## Author contributions

Conceptualization: A.M., J.H.M. and J.M.B. Investigation (numerical simulations) and formal analysis of the data: A.M. Formal analysis (analytical treatment): J.H.M. Supervision: J.M.B. Writing, original draft: A.M. Writing, review and editing: A.M., J.H.M. and J.M.B.

## Disclosure statement

The authors report there are no competing interests to declare.

## Declaration of generative AI use

The simulation, analysis and figure code was developed with the assistance of Claude (Anthropic), a generative AI tool, under the supervision of the authors. The authors reviewed the code, validated the Monte Carlo simulations against exact calculations and take full responsibility for the results and for the content of the manuscript.

## Data availability statement

The ATP and Grand Slam data analysed in this study were compiled by J. Sackmann and are available under a CC BY-NC-SA 4.0 licence from an archival copy of the original repositories (tennis-sackmann-archive, 2026). The simulation and analysis code (Python) is available at https://github.com/jmbuldu/crucial_point and will be made public upon publication, together with the processed data supporting the findings.

**Table 1. Winning more, or fewer, matches than the points imply.** The ten ATP players (1991–2025, at least 200 matches) with the largest and the ten with the smallest close match balance. Points and Matches: shares won (%), matches with the i.i.d. expectation in parentheses. Won with fewer points / lost with more points: paradoxical matches, with their i.i.d. expectations in parentheses. Balance: close match balance (matches) and *z*-score. *Significant after Benjamini–Hochberg correction (false discovery rate 5%). Close sets: share of sets decided by two games or fewer that were won (%). Asymmetry: set margin asymmetry (games). Pressure: pressure index (percentage points). Tiebreak: tiebreak overperformance (percentage points).

| Player | Matches played | Points (%) | Matches (%) | Won with fewer points | Lost with more points | Balance (*z*) | Close sets (%) | Asymmetry | Pressure | Tiebreak |
|---|---|---|---|---|---|---|---|---|---|---|
| **Largest balance** | | | | | | | | | | |
| Fabio Fognini | 774 | 49.4 | 51.2 (47.5) | 32 (14.8) | 10 (18.8) | +26.0* (+5.2) | 56.6 | +0.31 | +0.5 | +6.5 |
| John Isner | 775 | 50.6 | 61.7 (58.4) | 49 (21.5) | 32 (28.1) | +23.6* (+3.9) | 57.5 | +0.02 | −0.1 | +9.0 |
| Carlos Moyà | 835 | 51.1 | 63.6 (60.9) | 32 (19.0) | 14 (21.9) | +20.9* (+3.8) | 60.8 | −0.03 | +0.8 | +5.8 |
| Tommy Robredo | 849 | 50.8 | 60.0 (57.7) | 34 (19.6) | 13 (16.6) | +18.0* (+3.5) | 57.2 | +0.03 | +0.2 | +3.7 |
| Goran Ivanišević | 710 | 51.5 | 65.4 (62.9) | 25 (10.2) | 18 (20.8) | +17.6* (+3.6) | 63.0 | +0.03 | +1.8 | +6.2 |
| Feliciano López | 948 | 49.6 | 50.6 (48.6) | 38 (22.9) | 23 (25.0) | +17.1 (+2.9) | 53.0 | +0.20 | +1.0 | +4.1 |
| Kei Nishikori | 623 | 51.4 | 66.9 (63.9) | 20 (10.2) | 8 (14.8) | +16.6* (+3.9) | 61.0 | −0.01 | −0.6 | +5.9 |
| Yevgeny Kafelnikov | 826 | 51.5 | 66.5 (64.5) | 22 (14.2) | 13 (20.7) | +15.4* (+3.0) | 59.1 | −0.11 | +0.6 | +3.3 |
| Fernando González | 518 | 51.1 | 64.3 (61.7) | 22 (10.7) | 8 (9.8) | +13.1* (+3.3) | 63.2 | +0.02 | +1.3 | +7.3 |
| Jürgen Melzer | 615 | 49.7 | 51.2 (48.7) | 18 (13.0) | 8 (15.8) | +12.7 (+2.7) | 51.1 | +0.06 | +0.3 | −0.3 |
| **Smallest balance** | | | | | | | | | | |
| Roger Federer | 1,437 | 54.1 | 81.7 (82.9) | 9 (15.6) | 36 (25.6) | −17.0* (−3.0) | 69.2 | −0.54 | +0.2 | +7.7 |
| Fernando Verdasco | 962 | 50.9 | 55.0 (56.7) | 17 (18.6) | 33 (18.9) | −15.7* (−2.9) | 51.5 | −0.25 | +0.2 | −2.0 |
| Todd Martin | 591 | 51.9 | 63.6 (66.0) | 11 (13.3) | 26 (14.7) | −13.6* (−3.0) | 57.4 | −0.18 | +0.5 | +4.0 |
| Nikolay Davydenko | 759 | 51.8 | 61.0 (62.9) | 7 (14.8) | 16 (10.6) | −13.3* (−3.0) | 52.6 | −0.34 | +0.3 | −0.7 |
| Dominik Hrbatý | 611 | 50.5 | 51.6 (53.9) | 7 (12.8) | 20 (13.3) | −12.5 (−2.8) | 50.2 | −0.14 | 0.0 | −0.1 |
| Marin Čilić | 898 | 51.7 | 62.4 (64.1) | 10 (15.1) | 28 (20.9) | −12.2 (−2.3) | 54.0 | −0.25 | −0.4 | −0.9 |
| Max Mirnyi | 423 | 50.0 | 47.5 (50.2) | 8 (12.1) | 21 (13.3) | −11.7 (−2.7) | 49.0 | −0.03 | −0.2 | +2.4 |
| Sam Querrey | 676 | 50.6 | 53.6 (55.1) | 14 (15.6) | 28 (17.9) | −11.7 (−2.3) | 50.3 | −0.11 | −0.7 | −2.9 |
| Marc Kevin Goellner | 330 | 49.8 | 44.2 (47.9) | 4 (8.1) | 15 (7.5) | −11.6* (−3.4) | 45.1 | −0.12 | +0.1 | −9.9 |
| Jonathan Stark | 276 | 50.4 | 47.5 (51.6) | 4 (6.0) | 14 (4.5) | −11.5* (−4.1) | 46.3 | −0.17 | −1.2 | +0.5 |

# Supplementary material

*Beyond the Federer paradox in tennis*

## S1. Amplification of a constant point advantage

Supplementary Table S1 gives, for each scoring system, the probability of winning a match for selected values of the probability $p$ of winning a point, the slope of $P(p)$ at $p$ = 1/2 and the values of $p$ needed to win 80% and 90% of matches, together with the random walk estimate $(2N/\pi)^{1/2}$ of the slope (Eq. 4 of the main text).

**Supplementary Table S1.** Amplification of a constant probability $p$ of winning a point (i.i.d. model, exact calculation). $N$ is the mean number of points per match at $p$ = 1/2.

| Format | $P$(0.52) | $P$(0.54) | d$P$/d$p$ | $(2N/\pi)^{1/2}$ | $p$ for 80% | $p$ for 90% | $N$ |
|---|---|---|---|---|---|---|---|
| Tennis, best of 5 | 0.749 | 0.910 | 13.4 | 13.2 | 0.525 | 0.538 | 272 |
| Tennis, best of 3 | 0.705 | 0.859 | 10.7 | 10.3 | 0.531 | 0.548 | 165 |
| Volleyball | 0.697 | 0.848 | 10.2 | 10.6 | 0.533 | 0.550 | 177 |
| Table tennis, best of 7 | 0.665 | 0.803 | 8.5 | 8.3 | 0.540 | 0.560 | 109 |
| Badminton | 0.654 | 0.786 | 7.9 | 7.7 | 0.543 | 0.565 | 93 |
| Table tennis, best of 5 | 0.642 | 0.767 | 7.3 | 7.0 | 0.546 | 0.570 | 78 |

## S2. Scoring systems and crucial points

**Supplementary Table S2.** Scoring systems simulated and share of crucial points at $p_1 = p_2 = 1/2$ for each level, for crucial points of both players ("both") and of player A only ("own"). In the rally scoring sports the game and set levels coincide.

| Format | Set | Match | Game (both / own) | Set (both / own) | Match (both / own) |
|---|---|---|---|---|---|
| Tennis, best of 3 | 6 games, margin 2, tiebreak at 6–6 | Best of 3 | 29.3% / 14.6% | 3.0% / 1.6% | 1.3% / 0.7% |
| Tennis, best of 5 | As above | Best of 5 | 29.3% / 14.6% | 3.1% / 1.5% | 0.7% / 0.4% |
| Badminton | 21 points, margin 2, cap 30 | Best of 3 | n/a | 5.3% / 2.7% | 2.1% / 1.1% |
| Volleyball | 25 points, margin 2, fifth set to 15 | Best of 5 | n/a | 4.6% / 2.3% | 1.2% / 0.6% |
| Table tennis, best of 5 | 11 points, margin 2 | Best of 5 | n/a | 10.5% / 5.2% | 2.7% / 1.3% |
| Table tennis, best of 7 | 11 points, margin 2 | Best of 7 | n/a | 10.4% / 5.3% | 1.8% / 0.9% |

## S3. Existence and width of the paradox region

Supplementary Table S3 gives, for every format and level, the ratios of the partial derivatives of Eq. (5) of the main text at (1/2, 1/2), the opening angle of the paradox wedge and the range of $p_2$ for which a player with $p_1 = 0.49$ wins most of his matches while winning fewer than half of the points.

The region requires a hierarchy, with the points grouped into games or sets and these into a match. If a single cumulative score decides the winner, as in football, or in a single game to 21 points, the winner always has more points and no paradox is possible. In every format of Supplementary Table S3 the match is built from several games or sets, and the condition $P_2/P_1 > F_2/F_1$ holds at every level.

**Supplementary Table S3.** Existence and width of the paradox region, crucial points of both players (exact calculation).

| **Format** | **Level** | **$P_2/P_1$** | **$F_2/F_1$** | **Angle (°)** | **$p_2$ at $p_1 = 0.49$** |
|---|---|---|---|---|---|
| Tennis, best of 3 | Game | 0.441 | 0.415 | 1.24 | 0.524 |
| | Set | 0.041 | 0.031 | 0.57 | 0.70–0.75 |
| | Match | 0.020 | 0.012 | 0.45 | > 0.80 |
| Tennis, best of 5 | Game | 0.441 | 0.415 | 1.24 | 0.524 |
| | Set | 0.041 | 0.031 | 0.57 | 0.70–0.75 |
| | Match | 0.013 | 0.007 | 0.34 | > 0.80 |
| Badminton | Set | 0.074 | 0.057 | 1.00 | 0.63–0.66 |
| | Match | 0.036 | 0.022 | 0.79 | ≥ 0.72 |
| Volleyball | Set | 0.069 | 0.049 | 1.15 | 0.63–0.68 |
| | Match | 0.024 | 0.011 | 0.74 | ≥ 0.77 |
| Table tennis, best of 5 | Set | 0.147 | 0.119 | 1.58 | 0.57–0.58 |
| | Match | 0.045 | 0.026 | 1.04 | 0.69–0.78 |
| Table tennis, best of 7 | Set | 0.147 | 0.119 | 1.58 | 0.57–0.58 |
| | Match | 0.033 | 0.019 | 0.83 | ≥ 0.73 |

## S4. Validation of the Monte Carlo simulations

The exact calculation by dynamic programming reproduces the classical closed form probability of winning a tennis game. Monte Carlo estimates of the probability of winning a match were compared with the exact values for every format, level and definition of crucial point. Over the 51,115 points of the ($p_1$, $p_2$) grids with a nondegenerate probability, the standardised deviation $z = (P_{MC} - P_{exact})/\sigma$, with σ the binomial standard error, has mean 0.00 and 99.5% of the values satisfy $|z| < 3$. For the curves with a single $p$ (845 points) the standard deviation of $z$ is 0.76. The expected numbers of points won by each player also agree with the simulations (e.g. $E[N_A] = 81.58$ exact against 81.60 simulated in tennis over three sets with $p_1$

= 0.47, $p_2$ = 0.70 and crucial points at the game level). The grid excludes $p$ = 0 and $p$ = 1 because, for $p_1$ = 1 and $p_2$ = 0 (or the reverse), a deuce never ends.

## S5. Crucial points of one player only

In a variant of the model only the points that player A can close are crucial. He converts his own game, set and match points with probability $p_2$, while all the other points, including those in which the opponent can close a unit, are played with $p_1$. Supplementary Table S4 gives the existence criterion and the opening angle of the paradox wedge for this variant. The paradox region exists in all four sports and at every level, as in the main definition, and is about half as wide, because the number of crucial points is halved.

**Supplementary Table S4.** As Supplementary Table S3, for crucial points of player A only.

| **Format** | **Level** | $P_2/P_1$ | $F_2/F_1$ | **Angle (°)** |
|---|---|---|---|---|
| Tennis, best of 3 | Game | 0.181 | 0.172 | 0.48 |
| | Set | 0.020 | 0.015 | 0.28 |
| | Match | 0.010 | 0.006 | 0.22 |
| Tennis, best of 5 | Game | 0.181 | 0.172 | 0.48 |
| | Set | 0.020 | 0.015 | 0.28 |
| | Match | 0.007 | 0.004 | 0.17 |
| Badminton | Set | 0.036 | 0.028 | 0.47 |
| | Match | 0.018 | 0.011 | 0.39 |
| Volleyball | Set | 0.033 | 0.024 | 0.55 |
| | Match | 0.012 | 0.006 | 0.37 |
| Table tennis, best of 5 | Set | 0.068 | 0.056 | 0.70 |
| | Match | 0.022 | 0.013 | 0.50 |
| Table tennis, best of 7 | Set | 0.068 | 0.056 | 0.70 |
| | Match | 0.016 | 0.009 | 0.40 |

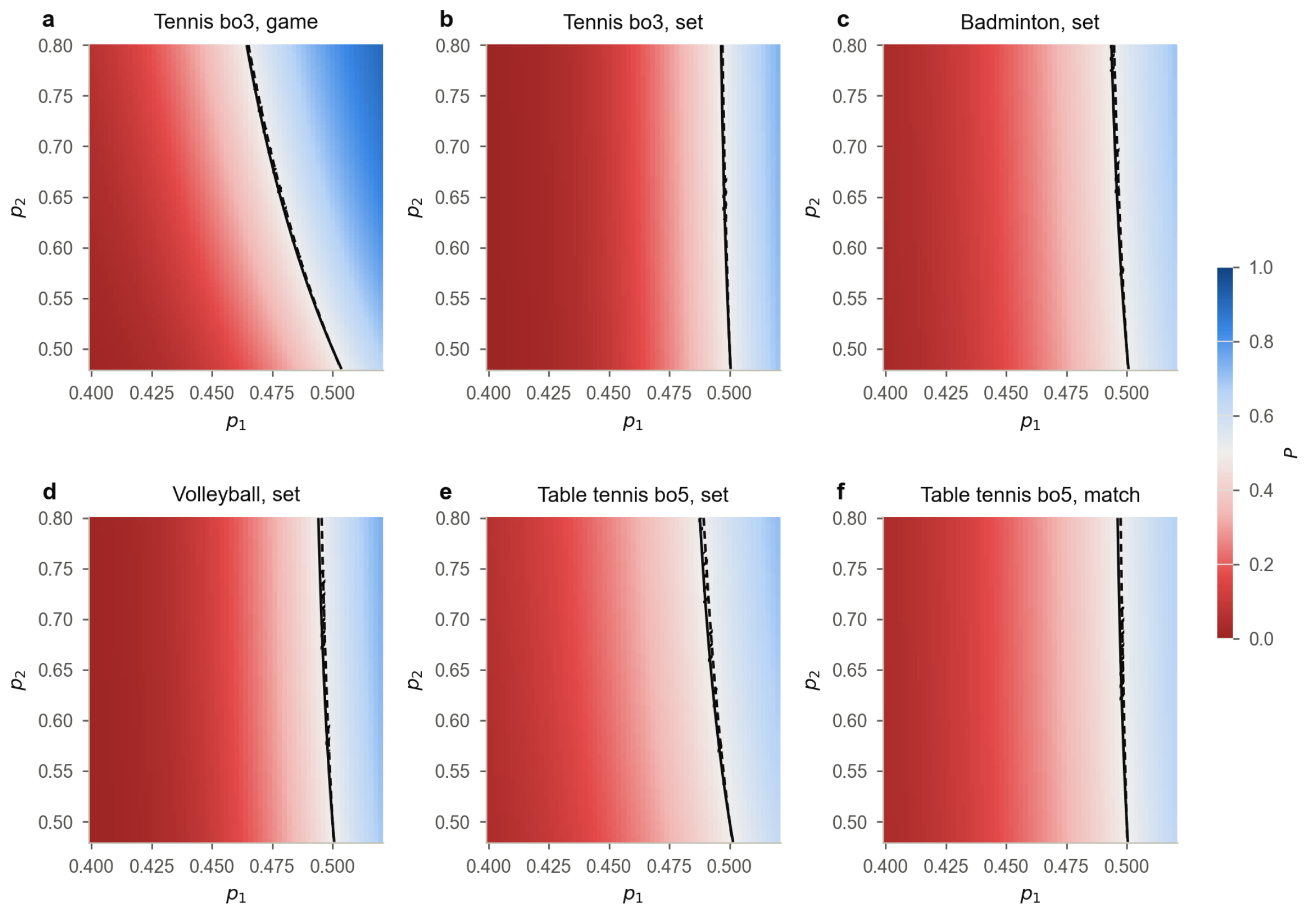

**Supplementary Figure S1 | Paradox regions when only the points that player A can close are crucial.** Probability $P$ of winning the match (colour) near (1/2, 1/2) for **a**, tennis (best of three) with crucial points at the game level, **b**, tennis at the set level, **c**, badminton, **d**, volleyball and **e**, table tennis (best of five) at the set level, and **f**, table tennis at the match level. Solid lines, $P = 1/2$. Dashed lines, $F = 1/2$. The paradox region lies between the two lines, to the left of the solid line and below the dashed one.

## S6. Single matches

The paradox region of the main text refers to averages over many matches. In single matches, a player can win while winning fewer than half of the points for any ($p_1$, $p_2$). Supplementary Figure S2 shows the probability of this event, obtained from 2,000 simulated matches per point of the grid.

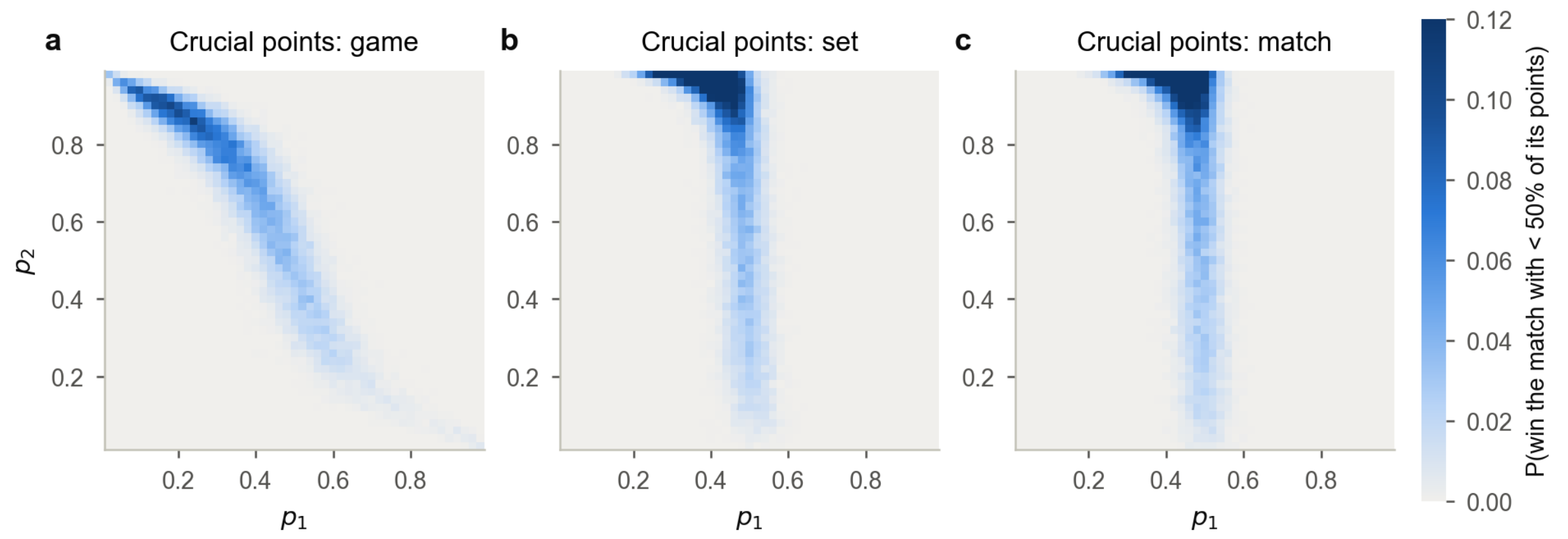

**Supplementary Figure S2 | Winning single matches with fewer points.** Probability that player A wins a tennis match over three sets while winning fewer than half of its points, as a function of $p_1$ and $p_2$, with crucial points at the level of **a**, the game, **b**, the set and **c**, the match (crucial points of both players, 2,000 Monte Carlo matches per grid point).

## S7. Data processing

**ATP matches.** We used the singles files of the main tour of the seasons 1991–2025 (the 2026 season was incomplete when the data were obtained). Matches were kept if both players had complete serve statistics (serve points, points won on the first and second serve, service games, break points faced and saved). Retirements, walkovers, defaults and unfinished matches were removed. The final data set contains 97,950 matches (161,226 records of a player in a match over three sets and 34,674 over five sets) and 2,559 players. The share of matches with statistics is 87–99% per season, the missing ones being mostly Davis Cup ties. For each player and match, the points won on serve are the points won on the first and second serve, and the points won on return are the opponent's serve points minus those he won. Break points on serve are the break points faced (won if saved) and break points on return are those faced by the opponent (won if not saved).

**Grand Slam point by point data.** Each file gives, for every point, the winner, the server, the set and game numbers and the score after the point. The score before a point is that after the previous point, reset at the end of each set. Men's singles matches were identified by their match number (1xxx, or MSxxx in the 2021 Australian Open). The rules of the final set are listed in Supplementary Table S5. Of 4,324 men's matches in the tournaments retained, 67 were removed because of inconsistent scores and 223 because they were not completed (28 for both reasons), leaving 4,062 matches. The Australian Open and Roland Garros files of 2018–2021 were excluded because the recorded game outcomes do not follow from the sequence of points (the score frequently goes backwards within a game), indicating that the order of points within matches is scrambled. In all other tournaments such inconsistencies affect fewer than 0.1% of points.

**Supplementary Table S5.** Rules of the fifth set used to reconstruct the score of Grand Slam men's singles matches.

| Tournament | 2011–2018 | 2019–2021 | 2022–2024 |
|---|---|---|---|
| Australian Open | Advantage set | 10-point tiebreak at 6–6 | 10-point tiebreak at 6–6 |
| Roland Garros | Advantage set | Advantage set | 10-point tiebreak at 6–6 |
| Wimbledon | Advantage set | 7-point tiebreak at 12–12 | 10-point tiebreak at 6–6 |
| US Open | 7-point tiebreak at 6–6 | 7-point tiebreak at 6–6 | 10-point tiebreak at 6–6 |

**Conversion curve.** The probability $\pi(f)$ of winning a match given the share of points won in it, $f$, was estimated from $4 \times 10^6$ simulated i.i.d. matches per format (best of three and best of five), with $p$ drawn uniformly in [0.3, 0.7], in bins of width 0.005 of $f$.

**Sliding windows.** Each window spans ten consecutive seasons (1991–2000 to 2016–2025) and includes the players with at least 50 matches in it. The standard deviation across players of the pressure index is corrected for binomial noise by subtracting the mean squared standard error from the observed variance.

## S8. Additional analyses of the ATP data

**Break points and pressure over three decades.** Supplementary Fig. S3 shows quantities computed in sliding windows of ten seasons. The amplification of the point advantage across players is stable (slope of the win rate against the share of points of about 8), break points have become rarer (from 9.1% to 8.3% of all points), the dispersion of the share of points across players peaked in the windows centred around 2010, and the correlation between the pressure index and match conversion grew from about 0.15 to 0.38, while the correlation between the pressure index and the overall level of the players vanished in the most recent windows.

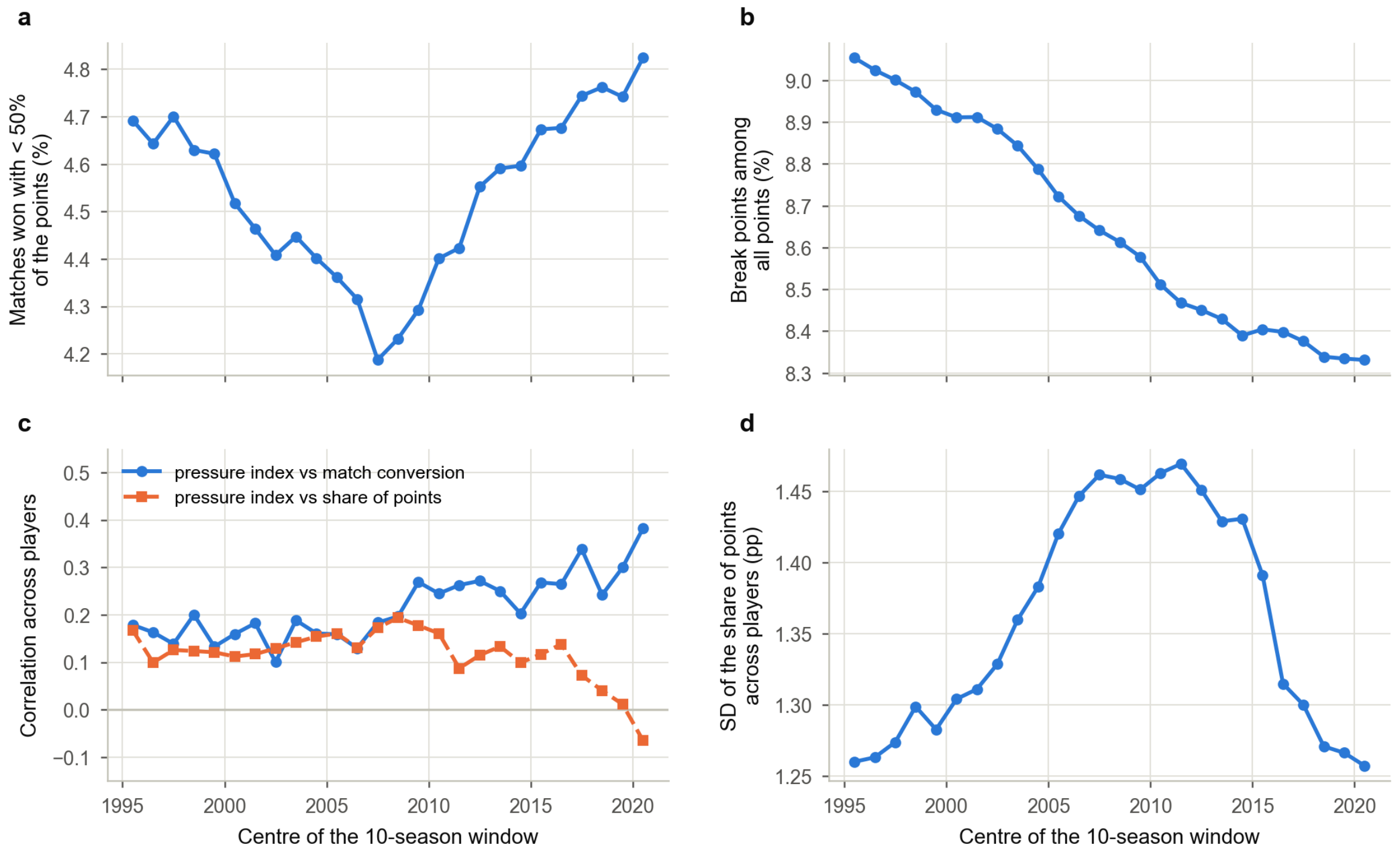

**Supplementary Figure S3 | Evolution of men's professional tennis, 1991–2025.** Sliding windows of ten seasons, plotted at the centre of each window. **a**, Share of matches won by the player who won fewer than half of the points. **b**, Share of break points among all points. **c**, Correlation across players (at least 50 matches in the window) between the pressure index and match conversion, and between

the pressure index and the share of points won. **d**, Standard deviation of the share of points won across players.

**Grand Slam players in the plane of the model.** Supplementary Fig. S4 places the 82 players with at least 30 Grand Slam matches (2011–2024) in the plane of the model, with $p_2$ measured on set and match points of both players. Most players lie close to the diagonal $p_2 = p_1$, within their binomial uncertainty, whereas Rafael Nadal and Novak Djokovic lie above it and Roger Federer slightly below it. Because winning a set requires winning the point that closes it, this measure within matches is partly mechanical and should be read together with the out of sample tests of the main text.

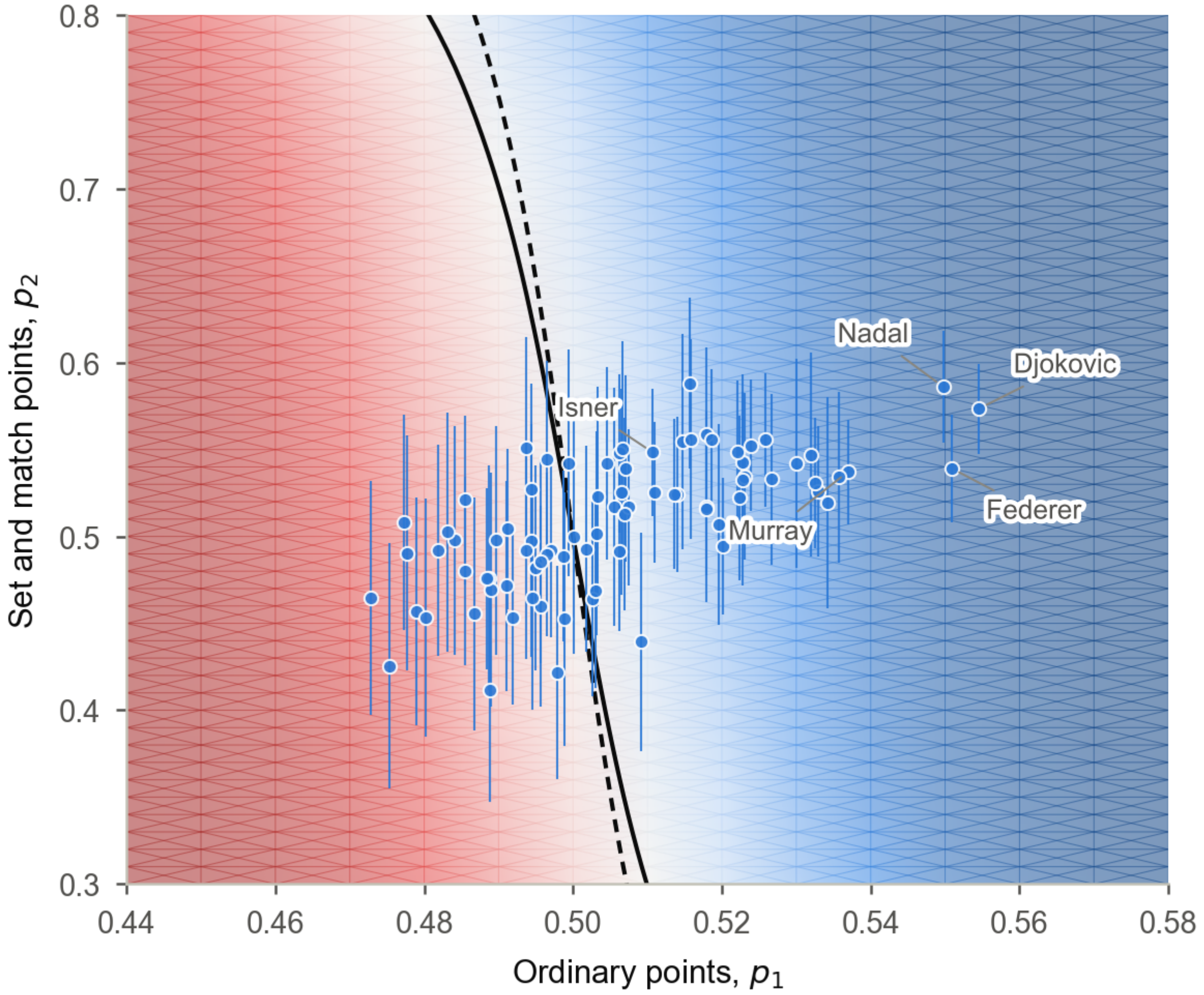

**Supplementary Figure S4 | Grand Slam players in the plane of ordinary and crucial points.** Probability of winning ordinary points, $p_1$, and set and match points, $p_2$, for players with at least 30 Grand Slam matches (2011–2024, error bars show 95% binomial intervals of $p_2$). The background colour gives the probability of winning a match over five sets in the model with crucial points at the set level. Solid line, $P = 1/2$. Dashed line, $F = 1/2$.

**Pressure ranking and location of the excess.** Supplementary Table S6 gives the players with the largest and smallest difference between their performance on crucial and on ordinary points (Fig. 5a of the main text), with the raw differences and those adjusted for the serve. Ranked by the raw difference, the bottom five would be Roger Federer (−4.1 pp), Carlos Alcaraz (−3.9), Andy Roddick (−3.7), Milos Raonic (−3.4) and Novak Djokovic (−3.0), all strong servers whose adjusted differences do not differ from zero. Supplementary Table S7 and Supplementary Fig. S5 break down the excess of matches won with fewer points by surface, tournament level and format. The excess is significant on every surface except carpet, where the number of matches is smallest, and in every tournament category, and it is largest in Davis Cup ties (34%, although with only 1,293 matches), in Grand Slams (22%), on grass (21%) and on hard courts (19%), and smallest on clay (13%).

**Supplementary Table S6.** Players with the largest and smallest difference between their performance on crucial (break) points and on ordinary points (ATP, 1991–2025, at least 200 matches), ordered by the difference adjusted for the serve shown in Fig. 5a of the main text. $p_1$: share of ordinary points (all except break points) won. $p_2$: share of break points won, on serve and on return. Raw difference: $p_2 - p_1$. Expected $p_2$: share of break points the player would win at his own serve and return rates on ordinary points. Adjusted difference: $p_2$ minus expected $p_2$ (the pressure index), with 95% bootstrap confidence intervals and Benjamini–Hochberg $q$-values. The raw difference is strongly influenced by playing style (its correlation with the share of serve points won is −0.40, against 0.15 for the adjusted difference), because strong servers play most of their break points on return, where points are harder to win.

| **Player** | **Matches** | **$p_1$ (%)** | **$p_2$ (%)** | **Raw $p_2 - p_1$ (pp)** | **Expected $p_2$ (%)** | **Adjusted difference (95% CI, pp)** | **$q$** |
|---|---|---|---|---|---|---|---|
| **Top 10** | | | | | | | |
| Nick Kyrgios | 297 | 51.2 | 51.8 | +0.7 | 49.1 | +2.7 (+1.0, +4.5) | 0.09 |
| Marcos Ondruska | 231 | 48.8 | 52.1 | +3.3 | 49.5 | +2.6 (+0.8, +4.3) | 0.13 |
| Martin Kližan | 261 | 49.0 | 51.8 | +2.9 | 49.6 | +2.2 (+0.7, +3.8) | 0.13 |
| Alexander Volkov | 379 | 49.7 | 52.1 | +2.3 | 50.0 | +2.1 (+0.8, +3.4) | 0.09 |
| Cameron Norrie | 409 | 50.5 | 52.2 | +1.7 | 50.2 | +2.0 (+0.7, +3.2) | 0.09 |
| Nicolás Massú | 409 | 49.6 | 51.9 | +2.3 | 50.0 | +1.9 (+0.8, +3.1) | 0.08 |
| Mikhail Kukushkin | 363 | 48.7 | 51.5 | +2.8 | 49.6 | +1.9 (+0.5, +3.3) | 0.22 |
| Jo-Wilfried Tsonga | 660 | 51.8 | 51.8 | 0.0 | 49.9 | +1.9 (+0.9, +2.9) | 0.08 |
| Goran Ivanišević | 710 | 51.5 | 52.1 | +0.5 | 50.2 | +1.8 (+0.8, +2.9) | 0.08 |
| Jannik Sinner | 392 | 53.8 | 52.9 | −0.9 | 51.1 | +1.8 (+0.4, +3.2) | 0.27 |
| **Bottom 10** | | | | | | | |
| Lukáš Lacko | 211 | 48.3 | 48.4 | +0.1 | 49.8 | −1.4 (−3.2, +0.5) | 0.61 |
| Christophe Rochus | 305 | 47.8 | 47.0 | −0.7 | 48.6 | −1.5 (−3.0, −0.2) | 0.41 |
| Daniel Gimeno Traver | 259 | 48.3 | 47.7 | −0.6 | 49.2 | −1.5 (−3.2, +0.1) | 0.57 |
| João Sousa | 447 | 49.1 | 48.1 | −1.0 | 49.7 | −1.6 (−2.8, −0.3) | 0.27 |
| Marco Cecchinato | 201 | 48.4 | 47.9 | −0.5 | 49.5 | −1.6 (−3.5, +0.3) | 0.61 |
| Michael Russell | 219 | 48.4 | 47.6 | −0.8 | 49.3 | −1.7 (−3.4, +0.2) | 0.54 |
| Jaume Munar | 269 | 50.1 | 48.2 | −1.9 | 50.1 | −1.9 (−3.5, −0.4) | 0.27 |
| Radu Albot | 232 | 48.7 | 47.1 | −1.6 | 49.1 | −2.0 (−3.6, −0.3) | 0.29 |
| Martin Damm Sr | 351 | 49.3 | 47.7 | −1.5 | 49.8 | −2.0 (−3.5, −0.6) | 0.14 |
| Hyung Taik Lee | 262 | 49.7 | 47.4 | −2.3 | 49.8 | −2.4 (−3.9, −0.9) | 0.09 |

**Supplementary Table S7.** Matches won with fewer than half of the points, observed and expected under the i.i.d. model given the share of points won in each match, by surface, tournament level and format (ATP, 1991–2025). $z$ from the Poisson-binomial variance.

| Group | Value | Matches | Observed (%) | Expected (%) | Ratio | $z$ | $P$ |
|---|---|---|---|---|---|---|---|
| Surface | Carpet | 5,702 | 4.58 | 4.35 | 1.05 | 1.0 | 0.339 |
| Surface | Clay | 31,554 | 4.31 | 3.83 | 1.13 | 5.0 | $< 10^{-4}$ |
| Surface | Grass | 10,021 | 5.15 | 4.24 | 1.21 | 5.1 | $< 10^{-4}$ |
| Surface | Hard | 49,549 | 4.83 | 4.06 | 1.19 | 9.8 | $< 10^{-4}$ |
| Level | Other main tour | 59,744 | 4.57 | 4.01 | 1.14 | 7.8 | $< 10^{-4}$ |
| Level | Davis Cup | 1,293 | 4.87 | 3.62 | 1.34 | 2.7 | 0.007 |
| Level | Grand Slam | 16,538 | 4.76 | 3.89 | 1.22 | 6.5 | $< 10^{-4}$ |
| Level | Masters 1000 | 18,617 | 4.93 | 4.21 | 1.17 | 5.5 | $< 10^{-4}$ |
| Best of | 3 | 79,642 | 4.65 | 4.04 | 1.15 | 9.8 | $< 10^{-4}$ |
| Best of | 5 | 17,184 | 4.80 | 3.92 | 1.23 | 6.8 | $< 10^{-4}$ |

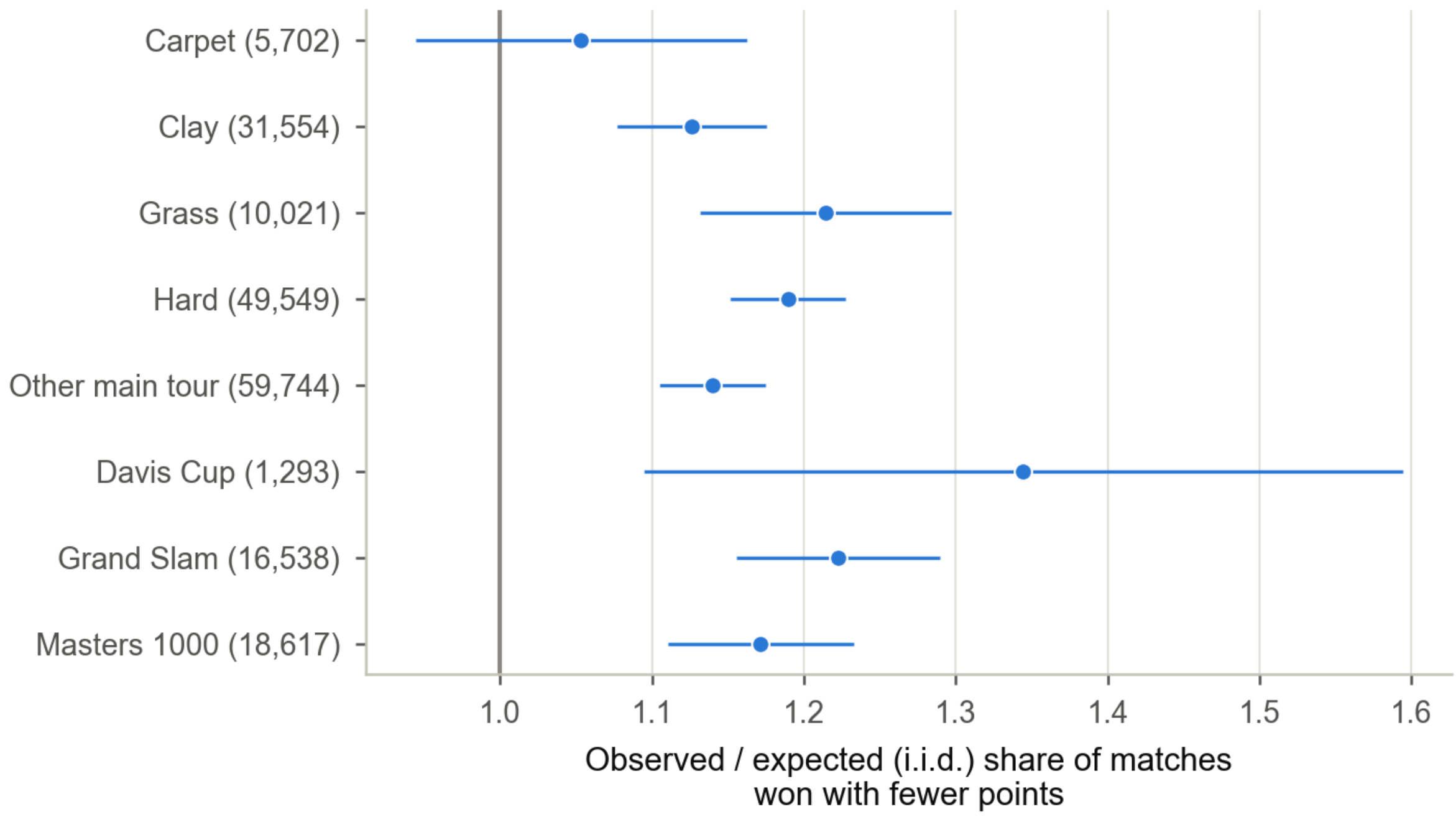

**Supplementary Figure S5 | Excess of matches won with fewer points by surface and tournament level.** Ratio between the observed and the expected number of matches under the i.i.d. model won with fewer than half of the points (95% confidence intervals). The number of matches is given in parentheses.

**Winning with fewer points, player by player.** Supplementary Table S8 lists the players with the largest positive and negative balance between matches won with fewer than half of the points and matches lost with more than half. Roger Federer has the most negative balance of the data set (9 won, 36 lost), the Federer paradox described by Lisi, Grigoletto and Canesso (2019), whereas Fabio Fognini, Tommy Robredo and Carlos Moyà have the most positive ones.

**Supplementary Table S8.** Players with the largest positive and negative balance of paradoxical matches (ATP, 1991–2025, at least 200 matches).

| Player | Matches | Won with < 50% of points | Lost with > 50% of points | Balance |
|---|---|---|---|---|
| **Most positive** | | | | |
| Fabio Fognini | 774 | 32 | 10 | +22 |
| Tommy Robredo | 849 | 34 | 13 | +21 |
| Carlos Moyà | 835 | 32 | 14 | +18 |
| John Isner | 775 | 49 | 32 | +17 |
| Guillermo García López | 634 | 21 | 6 | +15 |
| Feliciano López | 948 | 38 | 23 | +15 |
| Hendrik Dreekmann | 203 | 14 | 0 | +14 |
| Félix Mantilla | 510 | 18 | 4 | +14 |
| Fernando González | 518 | 22 | 8 | +14 |
| Jo-Wilfried Tsonga | 660 | 24 | 11 | +13 |
| **Most negative** | | | | |
| Roger Federer | 1,437 | 9 | 36 | −27 |
| Marin Čilić | 898 | 10 | 28 | −18 |
| Fernando Verdasco | 962 | 17 | 33 | −16 |
| Todd Martin | 591 | 11 | 26 | −15 |
| Sam Querrey | 676 | 14 | 28 | −14 |
| Juan Mónaco | 561 | 7 | 20 | −13 |
| Karen Khachanov | 518 | 7 | 20 | −13 |
| Roberto Bautista Agut | 693 | 10 | 23 | −13 |
| Max Mirnyi | 423 | 8 | 21 | −13 |
| Dominik Hrbatý | 611 | 7 | 20 | −13 |

**Paradoxical matches against their expectation.** Counting paradoxical matches favours players who rarely win more points than their opponents, and penalizes dominant players, who have many more chances to lose such a match than to win one. Supplementary Table S9 therefore compares, for selected champions, the observed numbers of matches lost despite winning more points and won despite winning fewer points with their expectations under the i.i.d. model given the share of points won in each match. The expectation of each match is the probability $\pi(f)$ of winning it with share of points $f$, and the variance is that of a Poisson-binomial distribution. The net balance, (won − lost) − (expected won − expected lost), is the number of close matches gained or lost with respect to independent points. Federer is the only champion with a negative balance (−17.0 matches, $z = -3.0$), whereas Sampras, Agassi, Murray, Nadal and, above all, Isner gained matches.

**Supplementary Table S9.** Paradoxical matches of selected players, observed and expected (in parentheses) under the i.i.d. model given the share of points won in each match (ATP, 1991–2025). *z*-scores from the Poisson-binomial variance. Net balance in matches.

| Player | Matches with more points | Lost (expected) | *z* | Matches with fewer points | Won (expected) | *z* | Net balance | *z* |
|---|---|---|---|---|---|---|---|---|
| Roger Federer | 1,197 | 36 (25.6) | +2.4 | 232 | 9 (15.6) | −1.9 | −17.0 | −3.0 |
| Rafael Nadal | 1,024 | 18 (17.1) | +0.2 | 217 | 21 (13.1) | +2.5 | +7.0 | +1.5 |
| Novak Djokovic | 1,101 | 15 (20.7) | −1.4 | 196 | 11 (15.2) | −1.3 | +1.5 | +0.3 |
| Pete Sampras | 623 | 19 (19.8) | −0.2 | 151 | 20 (9.6) | +3.8 | +11.2 | +2.4 |
| Andre Agassi | 629 | 12 (13.0) | −0.3 | 195 | 18 (10.5) | +2.6 | +8.4 | +2.0 |
| Andy Murray | 691 | 17 (19.0) | −0.5 | 255 | 17 (11.1) | +2.0 | +7.8 | +1.6 |
| Carlos Alcaraz | 268 | 4 (7.8) | −1.6 | 60 | 6 (4.9) | +0.6 | +4.9 | +1.6 |
| Jannik Sinner | 310 | 8 (6.6) | +0.6 | 78 | 9 (4.6) | +2.3 | +3.0 | +1.0 |
| John Isner | 451 | 32 (28.1) | +0.8 | 307 | 49 (21.5) | +6.9 | +23.6 | +3.9 |

**Tiebreaks.** The overperformance of a player in tiebreaks is the difference between the tiebreaks he won and those expected from the i.i.d. probability of winning a tiebreak to seven points at the share of points he won in each match. It is strongly correlated with match conversion ($r = 0.45$, 95% CI 0.36–0.53) but not with the pressure index on break points ($r = 0.09$, $P = 0.12$, Supplementary Fig. S6), so tiebreaks are a second, independent channel through which crucial points convert points into wins. Supplementary Table S10 lists the players with the largest overperformance among those with at least 100 tiebreaks.

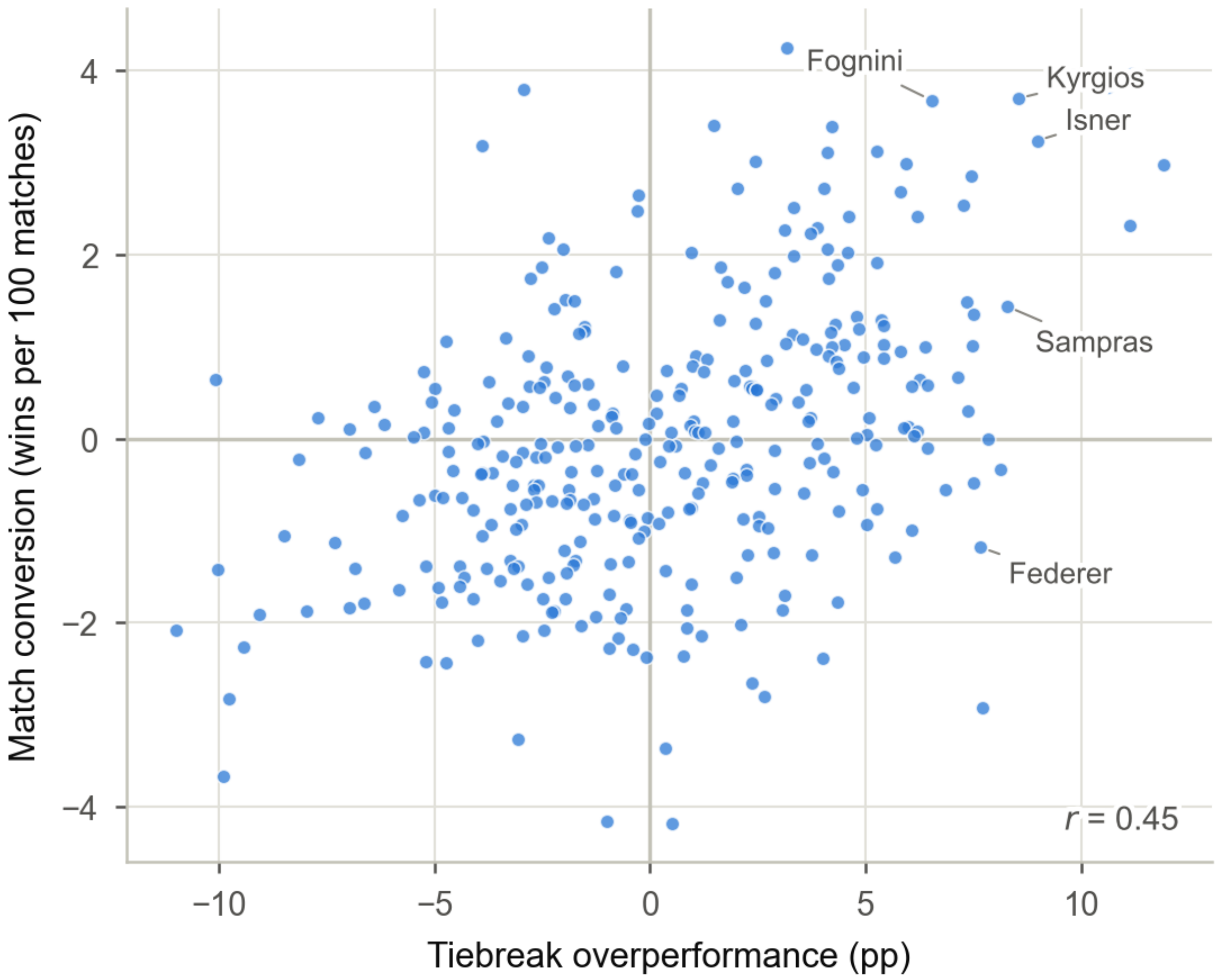

**Supplementary Figure S6 | Tiebreaks and match conversion.** Match conversion against tiebreak overperformance for the 328 ATP players with at least 200 matches (1991–2025).

**Supplementary Table S10.** Players with the largest tiebreak overperformance (at least 100 tiebreaks, ATP 1991–2025).

| Player | Tiebreaks | Won (%) | Overperformance (pp) | Match conversion (per 100 matches) |
|---|---|---|---|---|
| Jacco Eltingh | 130 | 60.8 | +11.1 | +2.3 |
| Nikoloz Basilashvili | 133 | 58.6 | +10.3 | +3.9 |
| John Isner | 808 | 60.5 | +9.0 | +3.2 |
| Nick Kyrgios | 221 | 59.3 | +8.5 | +3.7 |
| Pete Sampras | 413 | 63.9 | +8.3 | +1.4 |
| Novak Djokovic | 500 | 65.8 | +7.8 | 0.0 |
| Roger Federer | 670 | 65.1 | +7.7 | −1.2 |
| Holger Rune | 121 | 60.3 | +7.5 | +1.4 |
| Jannik Sinner | 158 | 63.9 | +7.5 | +1.0 |
| Yoshihito Nishioka | 103 | 55.3 | +7.4 | +2.9 |

## S9. Players beyond the Federer paradox

**The paradox over whole careers.** Supplementary Table S11 lists the 23 players with at least 200 ATP matches (1991–2025) who won more than half of their matches while winning less than half of the points, together with the share of matches they would be expected to win under the i.i.d. model given the share of points won in each match and their close match balance. For most of them the share of points is only slightly below one half. Fabio Fognini combines the lowest share of points of the group with the largest and most significant balance.

**Supplementary Table S11.** Players who won more than half of their matches with less than half of the points (ATP, 1991–2025, at least 200 matches), ordered by share of points won. Expected: share of matches expected under the i.i.d. model given the share of points of each match. Balance: close match balance (matches) and its $z$-score. $q$: Benjamini–Hochberg $q$-value.

| Player | Matches | Seasons | Points won (%) | Matches won (%) | Expected (%) | Balance | $z$ | $q$ |
|---|---|---|---|---|---|---|---|---|
| Bernard Tomic | 337 | 2009–2025 | 49.41 | 50.4 | 47.3 | +8.0 | +2.5 | 0.096 |
| Fabio Fognini | 774 | 2005–2025 | 49.45 | 51.2 | 47.5 | +26.0 | +5.2 | < 0.001 |
| Mikael Tillström | 210 | 1992–2000 | 49.52 | 50.5 | 46.7 | +7.0 | +2.5 | 0.097 |
| Feliciano López | 948 | 1998–2023 | 49.58 | 50.6 | 48.6 | +17.1 | +2.9 | 0.055 |
| Lucas Pouille | 271 | 2013–2025 | 49.63 | 50.9 | 47.8 | +9.0 | +2.9 | 0.049 |
| Reilly Opelka | 230 | 2016–2025 | 49.63 | 51.7 | 48.7 | +6.4 | +2.2 | 0.160 |
| Tallon Griekspoor | 226 | 2017–2025 | 49.66 | 51.3 | 48.6 | +7.1 | +2.2 | 0.144 |
| Jürgen Melzer | 615 | 1999–2018 | 49.70 | 51.2 | 48.7 | +12.7 | +2.7 | 0.062 |
| Alexander Bublik | 338 | 2016–2025 | 49.75 | 53.3 | 49.8 | +10.5 | +3.3 | 0.027 |
| Andrei Chesnokov | 303 | 1991–1999 | 49.77 | 51.2 | 50.2 | +2.4 | +0.7 | 0.736 |
| Cristian Garín | 242 | 2013–2025 | 49.83 | 51.7 | 50.2 | +2.6 | +0.8 | 0.708 |
| Jérôme Golmard | 257 | 1994–2004 | 49.84 | 52.5 | 51.6 | +0.8 | +0.3 | 0.929 |
| Todd Woodbridge | 397 | 1991–2001 | 49.86 | 51.1 | 49.5 | +4.6 | +1.4 | 0.446 |
| Jordi Arrese | 231 | 1991–1996 | 49.88 | 51.1 | 52.5 | −1.8 | −0.6 | 0.796 |
| Nicolás Lapentti | 536 | 1995–2010 | 49.91 | 50.2 | 50.7 | −3.4 | −0.8 | 0.732 |
| José Acasuso | 344 | 2001–2011 | 49.92 | 51.5 | 51.1 | +1.7 | +0.5 | 0.817 |
| Sebastián Báez | 208 | 2021–2025 | 49.94 | 50.5 | 51.4 | +0.1 | 0.0 | 0.991 |
| Dmitry Tursunov | 418 | 2001–2017 | 49.95 | 51.7 | 50.8 | +3.0 | +0.8 | 0.729 |
| Alexander Volkov | 379 | 1991–1997 | 49.95 | 54.9 | 52.2 | +9.5 | +2.7 | 0.074 |
| Jaime Yzaga | 225 | 1991–1996 | 49.95 | 53.3 | 53.7 | −0.7 | −0.3 | 0.929 |
| Mariano Zabaleta | 388 | 1995–2008 | 49.96 | 50.3 | 50.3 | −1.3 | −0.4 | 0.884 |
| Paul Haarhuis | 421 | 1991–1999 | 49.97 | 50.6 | 49.4 | +5.6 | +1.7 | 0.334 |
| Mariano Puerta | 227 | 1997–2005 | 49.98 | 53.3 | 50.3 | +6.3 | +2.6 | 0.077 |

**How Fognini loses his sets.** Supplementary Fig. S7 compares the distribution of the games won in lost sets for Fognini and three champions. A large share of the sets lost by Fognini were lopsided (18% ended 6–0 or 6–1), whereas the champions lost most of their sets by small margins. Points lost in such sets do not cost matches, which allows a player to win the majority of his matches with fewer than half of the points.

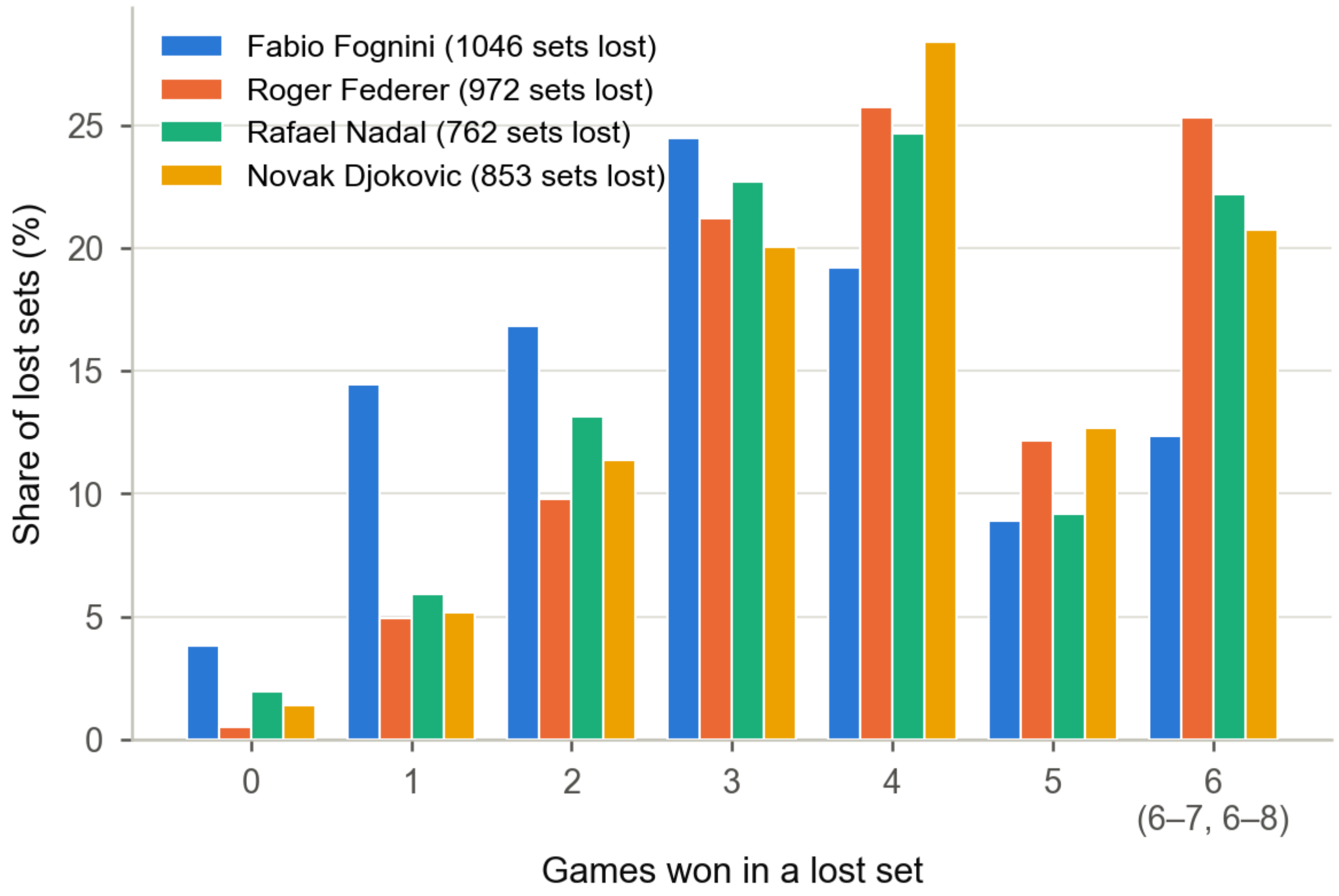

**Supplementary Figure S7 | Games won in lost sets.** Share of the sets lost by Fabio Fognini, Roger Federer, Rafael Nadal and Novak Djokovic in which they won 0, 1, …, 6 games (ATP, 1991–2025).

**Match conversion ranking.** Supplementary Table S12 ranks the players by match conversion, the average number of wins per 100 matches above those expected from the share of points won in each match, and includes a selection of champions. Match conversion and the close match balance are two forms of the same quantity. The balance equals the match conversion multiplied by the number of matches, computed without the matches in which both players won the same number of points (Methods). The balance expresses it in matches and separates the matches won with fewer points from those lost with more points.

**Supplementary Table S12.** Converting points into wins. The eight players with the highest and the five with the lowest match conversion, and selected champions (ATP, 1991–2025, at least 200 matches). Match conversion in wins per 100 matches. Pressure index and tiebreak overperformance in percentage points. Paradox W–L: matches won with fewer than half of the points and lost with more than half. Bootstrap 95% confidence intervals in parentheses. *Significant after Benjamini–Hochberg correction.

| Player | Matches | Points won (%) | Matches won (%) | Match conversion (95% CI) | Pressure index (95% CI) | Paradox W–L | Tiebreak |
|---|---|---|---|---|---|---|---|
| **Highest match conversion** | | | | | | | |
| Martin Kližan | 261 | 49.2 | 48.3 | +4.3* (+1.6, +7.0) | +2.2 (+0.7, +3.8) | 15–4 | +3.2 |
| Hendrik Dreekmann | 203 | 48.9 | 45.3 | +4.0 (+1.0, +7.0) | +0.3 (−1.6, +2.3) | 14–0 | +11.2 |
| Nikoloz Basilashvili | 311 | 48.8 | 45.7 | +3.9* (+2.1, +5.8) | −0.4 (−1.8, +1.1) | 11–1 | +10.3 |
| Mikael Tillström | 210 | 49.5 | 50.5 | +3.8 (+0.9, +6.8) | +0.1 (−1.5, +1.8) | 11–1 | +10.6 |
| Horacio Zeballos | 214 | 48.4 | 41.1 | +3.8 (+1.2, +6.8) | +1.5 (−0.7, +3.5) | 11–1 | −2.9 |
| Nick Kyrgios | 297 | 51.2 | 66.7 | +3.7* (+1.6, +5.9) | +2.7 (+1.0, +4.5) | 12–2 | +8.5 |
| Fabio Fognini | 774 | 49.4 | 51.2 | +3.7* (+2.3, +5.1) | +0.5 (−0.3, +1.4) | 32–10 | +6.5 |
| Alexander Bublik | 338 | 49.7 | 53.3 | +3.4* (+1.3, +5.6) | +1.6 (+0.2, +3.0) | 15–3 | +1.5 |
| **Lowest match conversion** | | | | | | | |
| Jonathan Stark | 276 | 50.4 | 47.5 | −4.2* (−6.9, −1.7) | −1.2 (−3.0, +0.4) | 4–14 | +0.5 |
| Jaume Munar | 269 | 49.9 | 42.4 | −4.2* (−6.4, −2.0) | −1.9 (−3.5, −0.4) | 2–10 | −1.0 |
| Marc Kevin Goellner | 330 | 49.8 | 44.2 | −3.7* (−5.9, −1.4) | +0.1 (−1.4, +1.6) | 4–15 | −9.9 |
| Brandon Nakashima | 233 | 50.9 | 53.2 | −3.4 (−6.4, −0.6) | −1.1 (−2.7, +0.6) | 4–12 | +0.4 |
| Andrei Cherkasov | 272 | 49.5 | 43.0 | −3.3 (−5.6, −1.0) | 0.0 (−1.4, +1.5) | 2–12 | −3.1 |
| **Selected champions** | | | | | | | |
| Roger Federer | 1,437 | 54.1 | 81.7 | −1.2 (−2.0, −0.3) | +0.2 (−0.6, +0.9) | 9–36 | +7.7 |
| Rafael Nadal | 1,248 | 54.4 | 82.8 | +0.8 (−0.1, +1.7) | +1.3 (+0.6, +2.1) | 21–18 | +4.4 |
| Novak Djokovic | 1,308 | 54.5 | 84.2 | 0.0 (−0.8, +0.8) | +0.6 (−0.1, +1.3) | 11–15 | +7.8 |
| Pete Sampras | 782 | 53.5 | 80.3 | +1.4 (0.0, +2.9) | +1.2 (+0.2, +2.2) | 20–19 | +8.3 |
| Andre Agassi | 830 | 53.3 | 77.0 | +1.1 (0.0, +2.3) | +0.8 (−0.1, +1.7) | 18–12 | +3.3 |
| Andy Murray | 949 | 52.7 | 73.0 | +0.9 (−0.1, +1.9) | −0.3 (−1.1, +0.5) | 17–17 | +5.4 |
| Carlos Alcaraz | 331 | 53.6 | 81.6 | +1.0 (−0.7, +2.7) | −0.8 (−2.1, +0.7) | 6–4 | +4.5 |
| Jannik Sinner | 392 | 53.7 | 80.1 | +1.0 (−0.7, +2.7) | +1.8 (+0.4, +3.2) | 9–8 | +7.5 |

## S10. Summary of statistical tests

**Supplementary Table S13.** Statistical tests on the ATP data quoted in the main text (i.i.d. benchmark, see Supplementary Section S11 for the benchmark that includes the serve).

| Quantity | Estimate | Test |
|---|---|---|
| Matches won with < 50% of points, 1991–2025 | 4.68% observed vs 4.02% expected (ratio 1.16) | $z = 11.7$, $P < 10^{-31}$ |
| Same, best of three / best of five | ratio 1.15 / 1.23 | $z = 9.8$ / 6.8 |
| Linear trend of the points paradox (logit per decade) | 0.011 (SE 0.015) | $P = 0.46$ |
| Quadratic term (U shape), minimum in 2007 | 0.059 (SE 0.016) | $P = 0.0003$ |
| Quadratic term controlled for closeness | 0.040 (SE 0.018) | $P = 0.03$ |
| Trend of the excess over the i.i.d. expectation | −0.04 pp per decade (SE 0.06) | $P = 0.51$ |
| Trend of the share of close matches | +0.19 pp per decade (SE 0.10) | $P = 0.07$ |
| Games paradox 1968–2025 (logit per decade) | 0.059 (SE 0.010) | $P < 10^{-7}$ |
| Games vs points paradox across seasons | $r = 0.58$ (95% CI 0.30–0.76) | $P = 0.0003$ |
| Out of sample pressure difference (logit per pp) | 0.076 (SE 0.011), 90,229 matches | $z = 7.2$, $P < 10^{-12}$ |
| Same, close matches ($0.48 < f < 0.52$) | 0.075 (SE 0.012), 17,801 matches | $z = 6.5$, $P < 10^{-10}$ |
| Pressure index (odd seasons) vs match conversion (even) | $r = 0.17$ | $P = 0.002$ |
| Match conversion, odd vs even seasons | $r = 0.35$ | $P < 10^{-10}$ |
| Win rate vs share of points (players) | $r = 0.97$ (0.965–0.977) | $P < 10^{-200}$ |
| Match conversion vs share of points | $r = -0.04$ (−0.15 to 0.07) | $P = 0.46$ |
| Pressure index vs share of points | $r = 0.21$ (0.10–0.31) | $P = 0.0002$ |
| Match conversion vs tiebreak overperformance | $r = 0.45$ (0.36–0.53) | $P < 10^{-16}$ |
| Pressure index vs tiebreak overperformance | $r = 0.09$ (−0.02 to 0.19) | $P = 0.12$ |
| Players with match conversion ≠ 0 (FDR 5%) | 20 of 328 | Benjamini–Hochberg |
| Players with pressure index ≠ 0 (FDR 5%) | 0 of 328 | Benjamini–Hochberg |
| Close match balance: players significant (FDR 5%) | 23 of 328 (16 positive, 7 negative) | Benjamini–Hochberg |
| Fognini: close match balance | +26.0 matches (32 won vs 14.8 and 10 lost vs 18.8 expected) | $z = 5.2$, $q < 0.001$ |
| Fognini: share of points below 1/2 | 61,624 of 124,623 (49.4%) | $z = -3.9$ |
| Federer: close match balance | −17.0 matches (36 lost vs 25.6 and 9 won vs 15.6 expected) | $z = -3.0$, $q = 0.046$ |
| Players winning most matches with < 50% of points | 23 of 328 | descriptive |
| Close match balance vs set margin asymmetry | $r = 0.33$ (0.23–0.42) | $P < 10^{-8}$ |
| Regression of the balance (standardised) | pressure 0.34, tiebreak 0.35, asymmetry 0.34, $R^2 = 0.37$ | all $P < 10^{-11}$ |

## S11. Robustness to the serve

The benchmark $\pi(f)$ used in the main text, the probability of winning a match given the share of points won in it, is computed with independent points of equal probability and therefore ignores the serve. In tennis the serve groups points into service games that are usually held, and this structure could by itself change the relation between points and wins. To test this, we simulated $6 \times 10^6$ matches over three sets and $3 \times 10^6$ over five sets in which each player wins the points he serves with his own probability, drawn uniformly in [0.48, 0.82] (the range of the ATP), the server alternating every game and, in tiebreaks, every two points. From these matches we estimated $\pi_s(f, g)$, the probability of winning given both the share of points $f$ and the serve dominance of the match, $g$, the sum of the shares of service points won by the two players (bins of 0.005 in $f$ and 0.05 in $g$, with exact ties of points kept apart and the estimate symmetrised). Each ATP match was then assigned the value of $\pi_s$ corresponding to its observed $f$ and $g$, and all the analyses that depend on the benchmark were repeated. For equal serve probabilities the simulator reproduces the exact i.i.d. results (Supplementary Section S4).

Supplementary Table S14 and Supplementary Fig. S8 summarise the results. For a given share of points, matches dominated by the serve are slightly less predictable (Supplementary Fig. S8a), so that more matches are expected to be won with fewer points. The excess of such matches over the expectation falls from 16% to 5%, but remains significant ($z = 3.9$). It falls to 3% in matches over three sets ($z = 2.3$) and remains 14% in matches over five sets ($z = 4.5$) (Supplementary Fig. S8b). Part of the excess reported with the i.i.d. benchmark is therefore due to the serve. In contrast, all the comparisons between players are essentially unchanged. The close match balance and the match conversion of the 328 players computed with both benchmarks are correlated with $r = 0.997$ (Supplementary Fig. S8c). Fabio Fognini (+25.3 matches, $z = 4.9$) and John Isner (+26.6, $z = 3.9$) remain at the top, the latter now first. Roger Federer remains second from the bottom (−14.6, $z = -2.4$, no longer significant after correction, $q = 0.12$). The out of sample effect of the pressure index is identical (0.077 per percentage point, $z = 7.2$). The three channels keep the same weights in the regression of the balance (0.34, 0.36 and 0.32, $R^2 = 0.37$). The difference between the two benchmarks is correlated with the serve strength of the players ($r = 0.33$), as expected, but is small compared with the differences between players. The number of players whose balance is significant after correction falls from 23 to 14.

**Supplementary Table S14.** Main empirical results with the i.i.d. benchmark (main text) and with the benchmark that includes the serve.

| Quantity | i.i.d. benchmark | Benchmark with serve |
|---|---|---|
| Matches won with < 50% of points: observed / expected | 4,530 / 3,892 (ratio 1.16, $z$ = 11.7) | 4,530 / 4,306 (ratio 1.05, $z$ = 3.9) |
| Same, best of three | ratio 1.15 ($z$ = 9.8) | ratio 1.03 ($z$ = 2.3) |
| Same, best of five | ratio 1.23 ($z$ = 6.8) | ratio 1.14 ($z$ = 4.5) |
| Trend of the paradox given the benchmark (logit per decade) | −0.014 ($P$ = 0.38) | −0.031 ($P$ = 0.06) |
| Out of sample pressure difference (logit per pp) | 0.076 ($z$ = 7.2) | 0.077 ($z$ = 7.2) |
| Correlation of player balances between benchmarks | n/a | $r$ = 0.997 |
| Players with significant balance (FDR 5%) | 23 | 14 |
| Fognini: close match balance | +26.0 ($z$ = 5.2) | +25.3 ($z$ = 4.9) |
| Isner: close match balance | +23.6 ($z$ = 3.9) | +26.6 ($z$ = 3.9) |
| Federer: close match balance | −17.0 ($z$ = −3.0, $q$ = 0.046) | −14.6 ($z$ = −2.4, $q$ = 0.12) |
| Regression weights (pressure, tiebreak, asymmetry) | 0.34, 0.35, 0.34, $R^2$ = 0.37 | 0.34, 0.36, 0.32, $R^2$ = 0.37 |

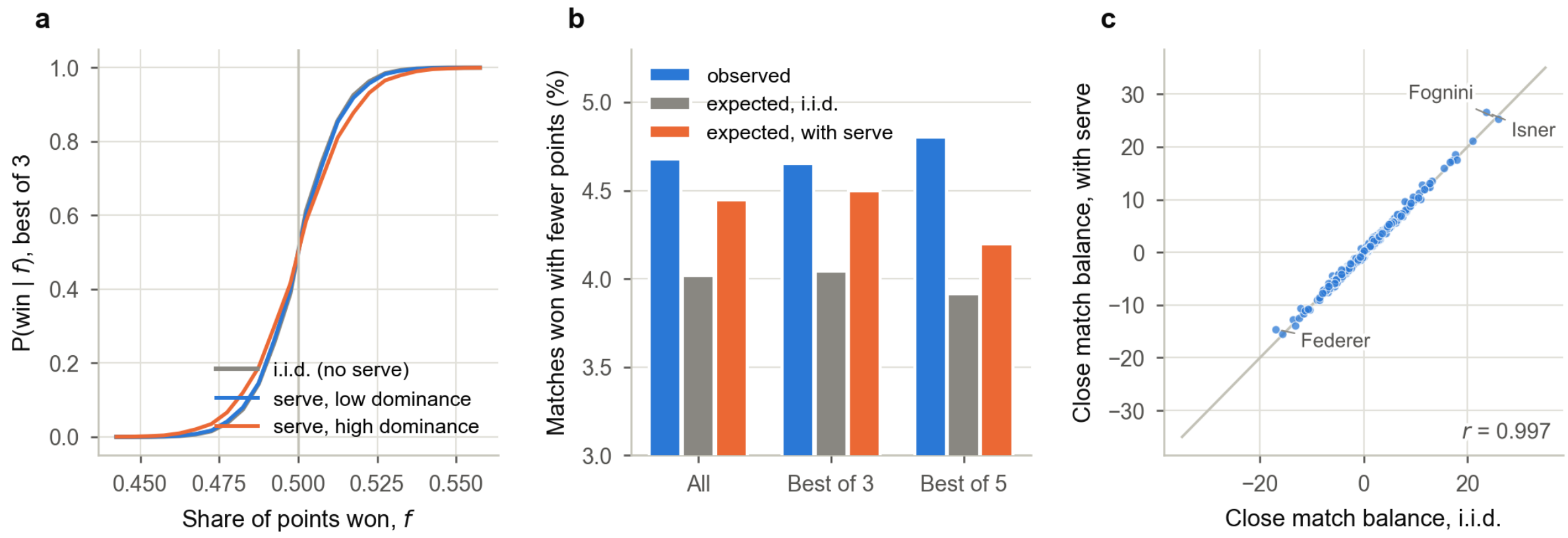

**Supplementary Figure S8 | Robustness to the serve. a**, Probability of winning a match over three sets as a function of the share of points won, for independent points without serve (grey) and for matches with the serve of low (blue, sum of the shares of service points won 1.15–1.20) and high (orange, 1.40–1.45) dominance. **b**, Observed share of ATP matches won with fewer than half of the points and its expectation with each benchmark. **c**, Close match balance of the 328 players with the two benchmarks. The line is the diagonal.

## S12. The model with the serve in every sport

The model of the main text gives every point the same probability, whoever serves. To test whether its conclusions survive the serve, we let player A win a point with probability $p + \sigma$ when he serves and $p - \sigma$ when his opponent serves, where $p = p_1$ for ordinary points and $p = p_2$ for crucial points, and $\sigma$ is the advantage of the server (negative when the receiver is favoured). This is the server model used in the literature on volleyball, badminton and table tennis, in which the probability of winning a rally depends on who serves (Paindaveine and Swan 2011, Dominicy et al. 2013, Ferrante and Fonseca 2014). For $p_1 = p_2 = 1/2$ the two players are still equivalent, so $P = F = 1/2$ and the analysis of Eqs. (5) and (6) of the main text applies unchanged. Each sport follows its own serve rules. In tennis the server changes every game and, in a tiebreak, every two points. In badminton and volleyball the winner of each rally serves the next one. In badminton the winner of a game serves first in the next one, whereas in volleyball the first serve alternates between sets and the deciding set starts with a toss. In table tennis the serve changes every two points and, from 10–10, every point, and the first server alternates between games. Who serves first in the match is decided by a toss.

We extended the exact calculation of the main text (Methods) to include the server in the score state. The deuce loop then has up to 12 states and is solved as a small linear system. For $\sigma = 0$ the calculation reproduces the results without serve to $10^{-13}$, and for $\sigma \neq 0$ it agrees with independent Monte Carlo simulations (20,000 to 400,000 matches per case) within sampling error for every sport. We scanned $\sigma$ from −0.20 to +0.20, a range that covers all four sports, and marked an indicative value for each of them. For $p = 1/2$, $\sigma$ is the share of rallies won by the server minus 1/2. For tennis the value is measured. The server won 62.7% of the points in our ATP data, so $\sigma = +0.13$. For the other sports we took indicative values from published rally statistics, which only illustrate the sign and the order of magnitude of the serve advantage. In volleyball the receiving team is clearly favoured. The serving team scored between 34% and 36% of the rallies of the four top teams of the Italian league in 2001–2012 (Ferrante and Fonseca 2014), and the receiving team won 69.3% of the rallies in men's matches and 64.9% in women's matches of the 2003 World Cup (Ogawa and Kurogo 2005). These figures give $\sigma$ between −0.14 and −0.19, and we used $\sigma \approx -0.16$. In badminton the server has no clear advantage. In the women's singles of the Rio 2016 Olympic Games the server won 49.6% of 1,052 rallies (Gómez-Ruano et al. 2020), so $\sigma = 0$ is compatible with the data and our value $\sigma \approx -0.03$ lies at the lower end of the 95% confidence interval of that proportion. In table tennis the server has a small advantage. In 140 matches of the Rio 2016 Olympic Games the server won on average 52.8% of the points in men's matches and 53.3% in women's matches (Gómez-Ruano et al. 2017), and we used $\sigma \approx +0.03$. Because the whole range of $\sigma$ is scanned, none of the conclusions below depends on these values.

**Amplification.** In the sports in which the server alternates in a fixed pattern (tennis and table tennis) the results are symmetric in $\sigma$. In badminton and volleyball, where the winner of a rally keeps the serve, they are not. At the indicative values the slope of $P(p)$ at $p = 1/2$ changes by −6% in tennis, +8% in volleyball and less than 1% in badminton and table tennis (Supplementary Table S15 and Supplementary Fig. S9a). The ranking of the six formats by amplification barely changes, and only volleyball overtakes tennis over three sets. In the rally scoring sports the serve only makes the points more or less predictable, and Eq. (4) of the main text remains accurate once the variance of a point, $1/4 - \sigma^2$ instead of 1/4, is taken into account, since the slope is within 5% of $(2N/\pi)^{1/2}(1 - 4\sigma^2)^{-1/2}$ for every σ (Supplementary Fig. S9c). In tennis the serve also groups the points into service games, which are usually held. Sets are then decided by a few breaks and by tiebreaks, and a point advantage is amplified less than the random walk estimate predicts (by 5% in matches over three sets and 7% over five sets at σ = 0.13).

**Paradox region.** The paradox region exists in all four sports, at every level and for every value of σ, because the condition $P_2/P_1 > F_2/F_1$ always holds. Its opening angle changes only moderately and, at the set and match levels, it widens (Supplementary Table S15 and Supplementary Fig. S9b). The largest change is in tennis at the set level, where the angle grows from 0.57° to 0.89° at σ = 0.13. With a strong serve, sets are decided by a few points, so the influence of a set point on the match grows ($P_2/P_1$ rises from 0.041 to 0.048) while its share of the points barely changes ($F_2/F_1$ from 0.031 to 0.032). At the game level, in contrast, the wedge of tennis narrows slightly (from 1.24° to 1.15°), because a game point on serve is likely to be won anyway. Over the realistic range of σ, all opening angles remain between 0.3° and 2°. The serve therefore changes the numbers by a few per cent but none of the conclusions of the model. Amplification is still set mainly by the length of the match, and a narrow paradox wedge still opens from (1/2, 1/2) in every sport.

**Supplementary Table S15.** The model with the serve (exact calculation). For each format, the indicative server advantage σ (measured for tennis, from published rally statistics for the other sports), the mean number of points per match $N$ at $p = 1/2$, the slope of $P(p)$ at $p = 1/2$, its ratio to the random walk estimate $(2N/\pi)^{1/2}(1 - 4\sigma^2)^{-1/2}$, and the opening angle of the paradox wedge for crucial points at each level (both players). Each entry gives the value without serve → with serve. In the rally scoring sports the game and set levels coincide.

| Format | σ | $N$ | Slope | Slope / random walk | Angle, game (°) | Angle, set (°) | Angle, match (°) |
|---|---|---|---|---|---|---|---|
| Tennis, best of 3 | +0.13 | 164.6 → 163.8 | 10.73 → 10.08 | 1.05 → 0.95 | 1.24 → 1.15 | 0.57 → 0.89 | 0.45 → 0.63 |
| Tennis, best of 5 | +0.13 | 271.6 → 270.3 | 13.42 → 12.60 | 1.02 → 0.93 | 1.24 → 1.16 | 0.57 → 0.87 | 0.34 → 0.46 |
| Badminton | −0.03 | 93.1 → 94.0 | 7.90 → 7.94 | 1.03 → 1.02 | n/a | 1.00 → 1.02 | 0.79 → 0.81 |
| Volleyball | −0.16 | 177.5 → 184.8 | 10.27 → 11.06 | 0.97 → 0.97 | n/a | 1.15 → 1.38 | 0.74 → 0.84 |
| Table tennis, best of 5 | +0.03 | 77.7 → 77.7 | 7.27 → 7.28 | 1.03 → 1.03 | n/a | 1.58 → 1.59 | 1.04 → 1.04 |

| **Format** | **σ** | ***N*** | **Slope** | **Slope / random walk** | **Angle, game (°)** | **Angle, set (°)** | **Angle, match (°)** |
|---|---|---|---|---|---|---|---|
| Table tennis, best of 7 | +0.03 | 109.4 → 109.5 | 8.48 → 8.50 | 1.02 → 1.02 | n/a | 1.58 → 1.59 | 0.83 → 0.83 |

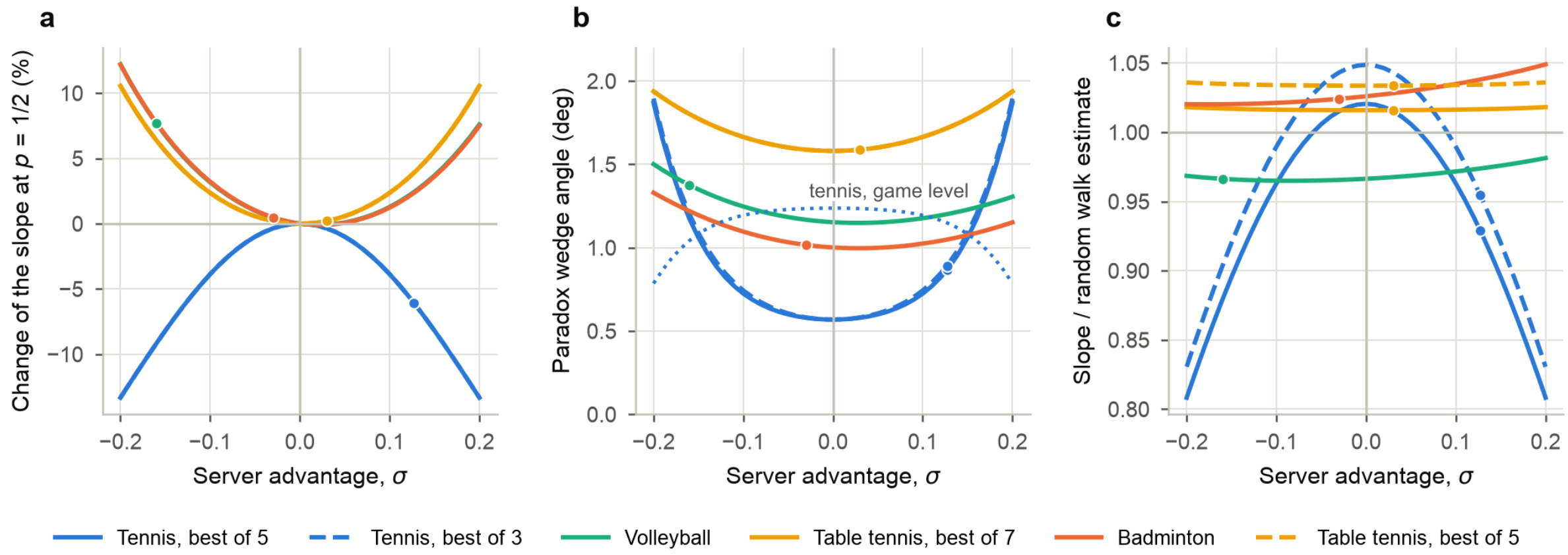

**Supplementary Figure S9 | The model with the serve in every sport.** Player A wins a point with probability $p + \sigma$ on his serve and $p - \sigma$ on his opponent's serve. **a**, Relative change of the slope of $P(p)$ at $p = 1/2$ as a function of the server advantage σ. **b**, Opening angle of the paradox wedge for crucial points at the set level (solid and dashed lines) and, for tennis, at the game level (dotted line). **c**, Ratio of the slope to the random walk estimate $(2N/\pi)^{1/2}(1 - 4\sigma^2)^{-1/2}$. Dots mark the indicative σ of each sport (measured for tennis, from published rally statistics for the other sports). The two formats of table tennis share the same curve in **b**. Exact calculation.

## S13. Simulations, exact calculation and derivation of Eq. (4)

This section gives the details of the calculations summarised in the Methods of the main text. A match is played between player A and player B. Player A wins an ordinary point with probability $p_1$ and a crucial point with probability $p_2$, and B wins them with probabilities $1 - p_1$ and $1 - p_2$. $N_A$ and $N_B$ are the numbers of points won by A and by B in a match, $P(p_1, p_2)$ is the probability that A wins the match and $F(p_1, p_2) = E[N_A]/(E[N_A] + E[N_B])$ is the share of points won by A over many matches.

**Monte Carlo simulations.** Matches are simulated point by point. The algorithm keeps the score of each match (points in the current game or tiebreak, games and sets) and, at each step, every unfinished match plays one point. The point is first classified as crucial or ordinary from the score (see Ordinary and crucial points), and A wins it if a uniform random number in [0, 1) is smaller than $p_2$ or $p_1$, respectively. The score is then updated with the rules of the sport. Many matches, each with its own $(p_1, p_2)$, are played simultaneously in vectorised form. For each match we record the winner, $N_A$, $N_B$ and the number of crucial points. Curves for a single $p$ were simulated with 20,000 matches per value of $p$, and the $(p_1, p_2)$ plane with 2,000 matches per point of a grid of step 0.02.

**Exact calculation.** Because the outcome of each point depends only on the current score, the score is a Markov chain and $P$ and $F$ can also be computed exactly. Let $s$ be a state of the score, $s_A$ and $s_B$ the states reached when A or B wins the next point, and $\rho(s)$ the probability that A wins that point, equal to $p_2$ if the point is crucial and to $p_1$ otherwise. The probability $V(s)$ that A wins the match from state $s$ obeys

$$V(s) = \rho(s)\, V(s_A) + [1 - \rho(s)]\, V(s_B), \qquad \text{(S1)}$$

with $V = 1$ in the states in which A has won the match and $V = 0$ in those in which B has won it. In the same way, the expected numbers of points that A and B will still win from state $s$, $a(s)$ and $b(s)$, obey

$$a(s) = \rho(s)[1 + a(s_A)] + [1 - \rho(s)]\, a(s_B), \quad b(s) = \rho(s)\, b(s_A) + [1 - \rho(s)][1 + b(s_B)], \qquad \text{(S2)}$$

with $a = b = 0$ at the end of the match. Then $P = V(s_0)$, $E[N_A] = a(s_0)$ and $E[N_B] = b(s_0)$, where $s_0$ is the score at the start of the match. Equations (S1) and (S2) are solved backwards from the end of the match (dynamic programming). Two details make the calculation simple. First, the match is a sequence of scoring units (games and tiebreaks in tennis, sets in the rally scoring sports), and during a unit the score at the higher levels does not change. Whether a point is crucial therefore depends only on the score of the unit and on which levels each player would close by winning the unit. For example, a game point in tennis is also a set point if winning that game would win the set. Each unit is solved once and the units are then combined. Second, a unit without a cap can last indefinitely after deuce. Its last states (tie, A one point ahead, B one point ahead) form a loop, and Eqs. (S1) and (S2) for these three states are a system of three linear equations, solved exactly. Badminton games, capped at 30 points, have no loop. The exact calculation reproduces the classical probability of winning a tennis game (Carter and Crews 1974), $p^4(1 + 4q + 10q^2) + 20p^5q^3/(1 - 2pq)$, with $q = 1 - p$, and the Monte Carlo estimates agree with the exact values within sampling error for every sport, level and definition of crucial point (Supplementary Section S4).

**Grids and derivatives.** $P$ and $F$ were computed exactly on a grid of step 0.01 in $[0.01, 0.99]^2$ and, around (1/2, 1/2), on finer grids of step 0.002 (Fig. 3) and of step 0.0005 to 0.001 (Fig. 2d–f). The partial derivatives of Eq. (5) of the main text were obtained by centred finite differences with $h = 10^{-4}$, for example $P_1 \approx [P(1/2 + h, 1/2) - P(1/2 - h, 1/2)]/(2h)$. The slope of Eq. (4) of the main text was obtained from the exact $P(p)$ with a centred difference of step 0.005 around $p = 1/2$.

**Derivation of Eq. (4) of the main text.** Consider the simplest contest between two players, in which $N$ points are played, with $N$ odd, and the player who wins more of them wins. If A wins each point with probability $p$, the number of points he wins follows a binomial distribution, and the probability that he wins the contest is

$$P_N(p) = \Sigma_{k=m+1}^{N} C(N, k)\, p^k q^{N-k}, \qquad (S3)$$

where $q = 1 - p$, $m = (N - 1)/2$ and $C(N, k)$ is the binomial coefficient. The derivative of $P_N$ has a simple meaning. Raising the probability of one particular point changes the outcome only if that point is decisive, that is, if the other $N - 1$ points are split evenly, $m$ for each player. This happens with probability $C(N - 1, m)(pq)^m$. All $N$ points play the same role, so

$$dP_N/dp = N\, C(N - 1, m)\, (pq)^m. \qquad (S4)$$

The same result is obtained by differentiating Eq. (S3) term by term, since all the terms but one cancel. Equation (S4) is a particular case of a general property. In any scoring system, $dP/dp$ is the expected sum, over the points played, of the probability that each point is decisive, that is, the expected total importance of the points of a match in the sense of Morris (1977). At $p = 1/2$, $pq = 1/4$ and Stirling's approximation (Feller 1968), $C(2m, m) \approx 4^m/(\pi m)^{1/2}$, gives

$$(dP_N/dp)_{p=1/2} = N\, C(N - 1, m)/2^{N-1} \approx N/(\pi m)^{1/2} \approx (2N/\pi)^{1/2}, \qquad (S5)$$

where the last step uses $m \approx N/2$. The approximation is very accurate. For $N = 79$ the exact slope is 7.11 and Eq. (S5) gives 7.09. The slope therefore grows as the square root of the number of points, because the probability that a given point is decisive decreases as $N^{-1/2}$ while the number of points grows as $N$.

The same result follows from the central limit theorem. The difference $D = N_A - N_B$ between the points won by the two players is a random walk of $N$ steps of $\pm 1$, with mean $N(2p - 1)$ and variance $4Npq$. A wins if $D > 0$, so for large $N$, $P_N \approx \Phi(N(2p - 1)/(4Npq)^{1/2})$, where $\Phi$ is the standard normal cumulative distribution function. Near $p = 1/2$, $4pq \approx 1$ and $P_N \approx \Phi(2(p - 1/2)N^{1/2})$. Since $\Phi'(0) = (2\pi)^{-1/2}$, the slope at $p = 1/2$ is $2N^{1/2}(2\pi)^{-1/2} = (2N/\pi)^{1/2}$.

Real scoring systems differ from this contest in two respects. First, the number of points is not fixed, and we take its mean at $p = 1/2$ (Supplementary Table S1). Second, the winner is not always the player who wins more points, and the hierarchy of games and sets redistributes importance among the points. Game, set and match points are much more important than the average point, and many other points are almost irrelevant. Equation (4) of the main text therefore assumes that the total importance of the points of a match is the same as in a contest of the same length decided by the total number of points. Figure 1b of the main text shows that this holds well. The exact slope is up to 5% higher than Eq. (4) in tennis, badminton and table tennis, and 3% lower in volleyball.

## Supplementary references